\documentclass[prd,eqsecnum,notitlepage,%linenumbers,                            %showpacs,%preprint,  
nofootinbib,superscriptaddress]
{revtex4-2}
\pdfoutput=1

\global\arraycolsep=2pt

\usepackage{stmaryrd}
\usepackage{amsmath}
\usepackage{amssymb}
\usepackage{graphicx}
\usepackage{textcomp}
\usepackage{calrsfs}
\usepackage{yfonts}
\usepackage{bm}
\usepackage{xcolor}
\newcommand{\ben}{\begin{equation*}}                                            
\newcommand{\een}{\end{equation*}}
\newcommand{\bean}{\begin{eqnarray*}}                                           
\newcommand{\eean}{\end{eqnarray*}}

\newcommand{\nn}{\nonumber}
\newcommand{\be}{\begin{equation}}                                             \DeclareMathOperator{\Si}{Si} 
\newcommand{\ee}{\end{equation}}
\newcommand{\bea}{\begin{eqnarray}}                                             
\newcommand{\eea}{\end{eqnarray}}

\DeclareMathOperator{\tr}{tr}

\DeclareMathOperator{\sgn}{sgn}
\DeclareMathOperator{\re}{Re}
\DeclareMathOperator{\im}{Im}

\begin{document}
\title{
Quantum Self-Propulsion of an Inhomogeneous Object out of Thermal
Equilibrium}
\author{Kimball A. Milton}
  \email{kmilton@ou.edu}
  \affiliation{H. L. Dodge Department of Physics and Astronomy,
University of Oklahoma, Norman, OK 73019, USA}
%\author{Prachi Parashar}                                           
%  \email{Prachi.Parashar@jalc.edu}                                   
%  \affiliation{John A. Logan College, Carterville, IL                
%62918, USA}                                                         
 
%\affiliation{Department of Energy and Process Engineering,            
%Norwegian University of Science and Technology, 7491 Trondheim, Norway}     

\author{Nima Pourtolami}
\email{nima.pourtolami@gmail.com}
\affiliation{National Bank of Canada, Montreal, Quebec H3B 4S9, Canada}

\author{Gerard Kennedy}\email{g.kennedy@soton.ac.uk}
\affiliation{School of Mathematical Sciences,
University of Southampton, Southampton, SO17 1BJ, UK}

%\author{Xin  Guo}
%\email{guoxinmike@ou.edu}
%\affiliation{H. L. Dodge Department of Physics and Astronomy, University of
%Oklahoma, Norman, OK 73019, USA}

\begin{abstract}In an earlier paper, we explored how quantum vacuum torque
can arise: a body or nanoparticle that is out of thermal
equilibrium with
its environment  %having a temperature $T'$ different from that the
%the blackbody background, $T$, 
experiences a spontaneous torque.
But this requires that the body be composed of nonreciprocal material,
which seems to necessitate the presence of an external influence, such
as a magnetic  field. Then the electric
polarizability of the particle has a real
part that is nonsymmetric.  This effect occurs to first order in the
polarizability.  To that order, no self-propulsive force can arise.  Here,
we consider second-order effects, and show that spontaneous forces
can arise in vacuum, without requiring exotic electromagnetic properties.
Thermal nonequilibrium is still necessary, 
but the electric
susceptibility of the  body need only be inhomogeneous.
We investigate four examples of such a body: 
a needle composed of distinct halves;
a sphere and a ball, each hemisphere being made of a different substance;
and a thin slab, each face of which is different.  The results found
are consistent with previous numerical investigations.
Here, we take into account the skin depth of  metal surfaces.
We also consider the frictional forces that would cause the body to acquire
a terminal velocity, which might be observable. 
More likely to be important is relaxation to
thermal equilibrium, which can still lead to 
a terminal velocity that might be experimentally verifiable.  
A general treatment of
such forces on a moving body, expressed in momentum space, is provided, 
which incorporates both propulsive and frictional forces.
The source of the propulsive force is the nonsymmetric
 pattern of radiation from different parts of the body, the higher
reflectivity of the metal portion playing a crucial role.
\end{abstract}

\date\today
\maketitle

\section{Introduction}
The subject of quantum vacuum forces on neutral particles or bodies has
a long history.  The classic analysis of Einstein and
Hopf \cite{EinsteinHopf} shows that a body moving 
nonrelativistically through vacuum experiences a frictional force opposing 
its motion that  depends on the temperature of the environment.  Dissipation
is required, which may arise either from intrinsic processes within the
body, producing an imaginary part of the electric susceptibility of the 
constituent material, or from the radiation field itself.  For a short 
representative list of papers on quantum vacuum friction, see
Refs.~\cite{Mrkt,dedkov2,Lach,pieplow,Volokitin0,vp}.

Much more work has been done on quantum electrodynamic friction near
dissipative surfaces; for example, see Refs.~\cite{teod, lev, pendry,
volokitin, dedkov, davies, barton, Zhao, Silveirinha, Pieplow15,
rev, Dedkov20, pan, Farias, Dedkov22, intravaia, marty,Wang,Brevik22}.  
In particular,
attention has been  directed to the force on a body or particle moving parallel
to a dielectric or conducting surface. This 
configuration will typically give rise to a much
larger frictional force, although even such friction has yet to be observed
in the laboratory.  These are nonequilibrium processes, but usually a
non-equilibrium steady state (NESS) configuration  has been considered,
where the moving object neither gains nor loses energy, but absorbs momentum
from the radiation field.  This automatically happens if the body is
nondissipative, with the radiation field being the only source of dissipation.
If the body is dissipative, satisfaction
of the NESS condition means that the temperature of the body bears a definite
ratio to that of the environment, that ratio depending of the velocity of
the body.  For low velocities, where NESS reduces to thermal equilibrium,
the ratio of temperatures tends to unity.

We recently revisited the subject of quantum vacuum friction, considering the
relativistic regime and nonequilibrium phenomena \cite{guo,guo2}. 
We proposed that the NESS temperature ratio might be an accessible signature
for experimentally observing such effects.  We considered both dipole
fluctuations and field fluctuations as sources for the dissipative effects.

But, even more remarkably, fluctuation phenomena may lead to spontaneous
torques and forces on stationary 
bodies out of equilibrium with their environment
\cite{reiche,krugeremig,Ott,maghrebi,Khandekar,khandekar0,fogedby,guofan,fan,
gelb,strekha,Ge}.
These effects, breaking time-reversal symmetry, typically require that the
bodies be constituted of nonreciprocal material, that is, that the electric 
susceptibility be non-Hermitian.\footnote{For much more on
nonreciprocal materials, see Ref.~\cite{tutorial}.}
  This may be achieved through the application
of an external magnetic field; see, for example, Ref.~\cite{guofan}.  
We explored such anomalous torques and forces in a recent paper \cite{QVT};
see also Ref.~\cite{kennedy}.
The effects considered there were first-order in the susceptibility or the
polarizability.  Although vacuum torques were found, no self-propulsive
forces could appear in first order, unless the body were adjacent to another
body, a dielectric or metal slab, for example.

To obtain a vacuum force for a body at rest requires not only nonequilibrium,
but also nonuniformity of the electric suseptibility,
 although  not necessarily nonreciprocity \cite{Reid,muller}.  
This requires going to second order in the susceptibility.  Even if the body
is spatially nonsymmetric but possesses uniform susceptibilty, 
no spontaneous force can arise, although it is claimed  that a
torque for such a body can appear if axial symmetry is broken, 
as for a pinwheel \cite{Reid}. 
(We will explore the quantum vacuum torque in a
subsequent paper.)  In this paper,
we demonstrate that a net force emerges if the susceptibility is nonuniform
across the object.  We analyze the general situation  for an arbitrary 
susceptibility (Sec.~\ref{sec1}), and apply the analysis 
to four particular configurations: an inhomogeneous needle 
(Sec.~\ref{sec:ihn}); an inhomogeneous
spherical shell (Sec.~\ref{sec:shell}); a ball with two halves made of 
different materials (a ``Janus ball'') \cite{Reid} (Sec.~\ref{sec:jb});  
and a thin plate consisting of metal on one side and  blackbody 
material\footnote{By blackbody material we mean a substance
whose mean electric
polarizability per unit surface area is such as to yield the
Planck spectrum and Stefan's law. See Sec.~\ref{sec:bbsurf}.} 
 on the other 
(Sec.~\ref{sec:bbsurf}). Of course, once a body is set in motion, a frictional
force will arise, resulting, eventually, in the body reaching a terminal
velocity if the friction is large enough; this is explored in 
Sec.~\ref{sec:fric}.
However, it is likely that thermal relaxation effects
will be much more important, and we show in Sec.~\ref{sec:thermal} that these
might still result in an experimentally measurable 
terminal velocity. An outlook is  provided in the
Conclusions.  Appendix \ref{appa}  provides a momentum-space treatment of the 
spontaneous force, and shows how second-order friction emerges naturally.
The thermal radiation from the body is
analyzed more fully in Appendix \ref{appb}; the nonsymmetric thermal radiation
pattern is responsible for the propulsive force. An explicit
example of a dilute dielectric body attached to a perfectly reflective 
mirror is given in Appendix \ref{appc}.

In this paper, we use natural units, with $\hbar=c=k_B=1$, and 
rationalized Heaviside-Lorentz electromagnetic units, with $\varepsilon-1=
\chi$ for the relation between the permittivity and the susceptibility.

\section{Self-propulsive force}
\label{sec1}
We follow the procedure detailed in Ref.~\cite{QVT,kennedy},
where we studied only nonreciprocal bodies and worked only to first order
in susceptibility,
and found spontaneous torques, but no vacuum self-propulsive force.

We started with the formula for the Lorentz force on a dielectric body
[Eq.~(7.1) of Ref.~\cite{QVT}]:\footnote{
The notation used in the second term here indicates, for
its $i$th component, $P_j\nabla_i E_j$.} 
\be
\mathbf{F}=\int(d\mathbf{r})\int \frac{d\omega}{2\pi}\frac{d\nu}{2\pi}
e^{-i(\omega+\nu)t}
\left\{-\left(1+\frac\omega\nu\right)
\left[\bm{\nabla}\cdot \mathbf{P}(\mathbf{r};\omega)\right]
\mathbf{E}(\mathbf{r};\nu)
-\frac\omega\nu
\mathbf{P}(\mathbf{r};\omega)\cdot(\bm{\nabla})\cdot
 \mathbf{E}(\mathbf{r};\nu)\right\}.\label{Lorentzforce}
\ee
Here, the first term will not contribute to the corresponding quantum
force, because the expectation value will make the sum of the frequencies
vanish, $\omega+\nu=0$. [See Eq.~(\ref{fdtset}) below.]
Then, as in the application of the  principle of virtual work
employed in Refs.~\cite{guo,guo2}, we see that the force density
may be thought of as the negative gradient of the interaction 
(free) energy density, 
\be
\mathfrak{f}=-\mathbf{P(r},t)\cdot \mathbf{E(r},t),
\label{freeen}
\ee
where the gradient acts only on the second, $\mathbf{E}$, factor.
 In Ref.~\cite{QVT} we expanded, to first order,
 the electric field  in terms of the electric polarization,
\begin{subequations}
\label{eppe}
\begin{equation}                           
\mathbf{E}^{(1)}(\mathbf{r};\nu)=\int (d\mathbf{r'})\,  
\bm{\Gamma}(\mathbf{r,r'};\nu)\cdot 
\mathbf{P}(\mathbf{r}';\nu),\label{EofP}  
\end{equation}
where $\bm{\Gamma}$ is the retarded electromagnetic Green's dyadic,
or we expanded the polarization in terms of  the electric field,
\begin{equation}   
\mathbf{P}^{(1)}(\mathbf{r};\omega)
%=\int (d\mathbf{r})\,\delta(\mathbf{r-r'})   
%\bm{\chi}(\mathbf{r};\omega)\cdot \mathbf{E(r';\omega)}=           
=\bm{\chi}(\mathbf{r};\omega)\cdot\mathbf{E}(\mathbf{r};\omega),\label{PofE} 
\end{equation}
\end{subequations}
where $\bm{\chi}=\bm{\varepsilon}-\bm{1}$ 
is the (local) electric susceptibility of the body.

In Ref.~\cite{QVT}, we found a torque on a nonreciprocal body.
We did so by applying the fluctuation-dissipation theorem (FDT)
to evaluate field products:
\begin{subequations}
\label{fdtset}
\begin{eqnarray}  
\langle \mathcal{S} E_i(\mathbf{r};\omega)E_j(\mathbf{r}';\nu)\rangle&=&
2\pi\delta(\omega+\nu)(\Im\bm{\Gamma})_{ij}(\mathbf{r,r'};\omega)\coth    
\frac{\beta\omega}2,\quad \beta=\frac1T,                           
 \label{eefdt}
\\                                            
\langle \mathcal{S} P_i(\mathbf{r};\omega)P_j(\mathbf{r}';\nu)          
\rangle&=&2\pi\delta(\omega+\nu)\delta(\mathbf{r-r'})            
(\Im\bm{\chi})_{ij}(\mathbf{r};\omega)\coth              
\frac{\beta'\omega}2,\quad \beta'=\frac1{T'},\label{ppfdt}     
\end{eqnarray}
\end{subequations}
where $\mathcal{S}$ denotes a symmetrized product, 
and $\Im$ signifies the anti-Hermitian part,\footnote{Fpr  a reciprocal susceptibility, the anti-Hermitian
part reduces to the usual imaginary part.}
to obtain first-order effects. Here, $T$ is the temperature of the blackbody
background, while $T'$ is that of the body.
 But there is, in vacuum, no first-order
force, because the gradient of the
Green's dyadic is then evaluated symmetrically 
at coincident points.

In the current paper, we consider second-order effects.
%but consider bodies are made of reciprocal material.
Iterating Eqs.~(\ref{EofP}) and (\ref{PofE}) results in
\begin{subequations}
\bea
\mathbf{E}^{(2)}(\mathbf{r};\nu)
&=&\int(d\mathbf{r'})\,\bm{\Gamma}(\mathbf{r,r'};\nu)
\cdot\bm{\chi}(\mathbf{r}';\nu)\cdot\mathbf{E}(\mathbf{r'};\nu),\\
\mathbf{E}^{(3)}(\mathbf{r};\nu)
&=&\int(d\mathbf{r'})(d\mathbf{r''})\,\bm{\Gamma}(\mathbf{r,r'};\nu)
\cdot\bm{\chi}(\mathbf{r}';\nu)\cdot\bm{\Gamma}(\mathbf{r',r''};\nu)\cdot
\mathbf{P(r'';\nu)},\\
\mathbf{P}^{(2)}(\mathbf{r};\omega)
&=&\int(d\mathbf{r'})\,\bm{\chi}(\mathbf{r};\omega)
\cdot\bm{\Gamma}(\mathbf{r,r'};\omega)\cdot\mathbf{P}(\mathbf{r'};\omega),\\
\mathbf{P}^{(3)}(\mathbf{r};\omega)
&=&\int(d\mathbf{r'})\,\bm{\chi}(\mathbf{r};\omega)\cdot
\bm{\Gamma}(\mathbf{r,r'};\omega)\cdot\bm{\chi}(\mathbf{r'};\omega)
\cdot\mathbf{E}(\mathbf{r'};\omega),
\eea
\end{subequations}
the order referring to the number of 
(generalized) susceptibility factors, 
either $\bm{\chi}$  or $\bm{\Gamma}$.
The second-order terms in the interaction energy 
(fourth order in susceptibilities) are obtained  by
 using (1) $\mathbf{P}^{(0)}$ and $\mathbf{E}^{(3)}$,
 (2) $\mathbf{P}^{(2)}$ and $\mathbf{E}^{(1)}$,  both of which give
quadratic structures in $\mathbf{P}$, while quadratic structures in 
$\mathbf{E}$
emerge by using (3) $\mathbf{P}^{(1)}$ and $\mathbf{E}^{(2)}$, and 
(4) $\mathbf{P}^{(3)}$ and $\mathbf{E}^{(0)}$.
The FDT (\ref{ppfdt}) is applied for the former, while
 the FDT  (\ref{eefdt}) is applied for the latter.
The resulting terms in the force will be denoted
$\mathbf{F}^{(0,3)}$, $\mathbf{F}^{(2,1)}$, $\mathbf{F}^{(1,2)}$, 
and $\mathbf{F}^{(3,0)}$, respectively,
in an obvious notation.

The EE fluctuations give rise to a force contribution in the $z$ direction of
\begin{subequations}
\bea
F_z^{\rm EE}&=&\int(d\mathbf{r})(d\mathbf{r'}) \frac{d\omega}{2\pi}
\chi_{ij}^{\vphantom{q}}(\mathbf{r};\omega)
\bigg[\nabla_z\Gamma_{ik}(\mathbf{r,r'};-\omega)
\chi^{\vphantom{q}}_{kl}(\mathbf{r}';-\omega)
(\Im \Gamma)_{jl}(\mathbf{r,r'};\omega)\nonumber\\
&&\quad\mbox{}+\Gamma_{jk}(\mathbf{r,r'};\omega)
\chi_{kl}^{\vphantom{q}}(\mathbf{r'};\omega)\nabla_z
(\Im\Gamma)_{li}(\mathbf{r',r};\omega)\bigg]\coth\frac{\beta\omega}2,
\label{FEEz}
\eea
while the PP fluctuations give
\bea
F_z^{\rm PP}&=&\int(d\mathbf{r})(d\mathbf{r'}) \frac{d\omega}{2\pi}
\nabla_z\Gamma_{jk}(\mathbf{r,r'};-\omega)
\bigg[\chi_{kl}^{\vphantom{q}}(\mathbf{r'};
-\omega)\Gamma_{lm}(\mathbf{r',r};-\omega)
(\Im \chi)_{jm}(\mathbf{r};\omega)\nonumber\\
&&\mbox{}+(\Im\chi)_{mk}(\mathbf{r'};\omega)
\Gamma_{lm}(\mathbf{r,r'};\omega)
\chi_{jl}(\mathbf{r};\omega)\bigg]\coth\frac{\beta'\omega}2.
\label{FPPz}
\eea
\end{subequations}
Equation (\ref{FEEz}) displays the (1,2) and (3,0) contributions, respectively,
while Eq.~(\ref{FPPz}) displays  the (0,3) amd (2,1) contributions,
respectively.

Henceforth, we restrict attention to reciprocal bodies.
The general result is particularly simple if the susceptibility is isotropic,
\be
\chi_{ij}^{\vphantom{q}}(\mathbf{r}; \omega)=\delta_{ij}
\chi(\mathbf{r};\omega),
\ee
and the background is not only reciprocal,
\be
\Gamma_{ij}(\mathbf{r,r'};\omega)=\Gamma_{ji}(\mathbf{r',r};\omega),
\ee
but also homogeneous, in which case the Green's dyadic is  a function only
of $\mathbf{R=r-r'}$.
We then use symmetry under $\mathbf{r}\leftrightarrow
\mathbf{r'}$ and also $\omega\to-\omega$,
noting that the imaginary part of the susceptibility is odd in
$\omega$, to derive
 the quantum vacuum force in the $z$-direction, which is, including the
contributions from EE and PP fluctuations,
\be
F_z=\int(d\mathbf{r})(d\mathbf{r'})\int_{-\infty}^\infty \frac{d\omega}{2\pi}
X(\mathbf{r,r'};\omega)\im \Gamma_{ji}(\mathbf{r'-r};\omega)
\nabla_z\im \Gamma_{ij}(\mathbf{r-r'};\omega)
\left[\coth\frac{\beta\omega}2-\coth\frac{\beta'\omega}2\right],
\label{qvf}
\ee
where the second-order susceptibility product is
\be
X(\mathbf{r,r'};\omega)
=\im \left[\chi(\mathbf{r};\omega)
\chi(\mathbf{r'};\omega)^*\right]
=\im\chi(\mathbf{r};\omega)
\re\chi(\mathbf{r'};\omega)-\re\chi(\mathbf{r};\omega)
\im\chi(\mathbf{r'};\omega).\label{secordsus}
\ee
(Note that only the first term of Eq~(\ref{FEEz}),  
$F^{(1,2)}$, contributes in the isotropic case, because the integrand
of the second term
is antisymmetric under $\mathbf{r}\leftrightarrow\mathbf{r}'$.)
It is clear that
if the susceptibility of the body is  homogeneous, 
so $\chi$ is independent of $\mathbf{r}$
within the body, the force is zero, since $X=0$ then.
%the resulting integrand
%changes sign under $\mathbf{r\leftrightarrow r'}$, because the
%Green's function is only a function of $|\mathbf{r-r'}|$.  
So, we need an inhomogeneous body
in order to generate a spontaneous quantum vacuum
force.

\begin{figure}
\includegraphics{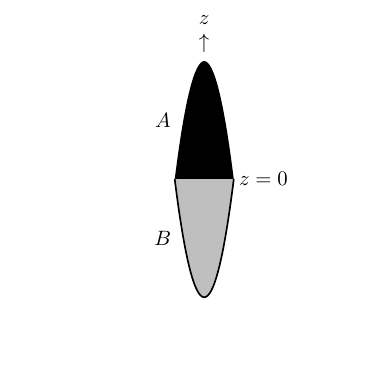}
\caption{Cartoon of the inhomogeneous bodies considered in this paper.
The upper part, $A$, will typically be a (nondispersive) insulator, while
the lower part, $B$, will be a Drude-type metal.  The figure is intended to
illustrate a solid body, rotationally invariant about the $z$-axis.  Because
of this symmetry, the force must be either parallel or antiparallel to the 
$z$-axis.}
\label{cartoon}
\end{figure}

The next simplest
 possibility is to imagine the object as consisting
of two separate parts, labeled $A$ and $B$, in each of which $\chi$ is
spatially constant, as illustrated in Fig.~\ref{cartoon}.
  Then the formula for the force reduces to 
\begin{subequations}
\be
F_z=8\int_0^\infty
\frac{d\omega}{2\pi}X_{AB}^{\vphantom{q}}
(\omega)
\left[\frac1{e^{\beta\omega}-1}-\frac1{e^{\beta'\omega}-1}\right]
I_{AB}(\omega),
\label{forceAB}
\ee
where
\be
I_{AB}(\omega)=
\int_A(d\mathbf{r})\int_B(d\mathbf{r'})\,
\frac12\nabla_z\left[\im \Gamma_{ji}(\mathbf{r'-r};\omega)
\im \Gamma_{ij}(\mathbf{r-r'};\omega)\right],\label{IAB}
\ee
which encodes the geometry of the configuration,
and
\be
X_{AB}^{\vphantom{q}}(\omega)=\im\chi_A^{\vphantom{q}}(\omega)
\re\chi_B^{\vphantom{q}}(\omega)
-\re\chi_A^{\vphantom{q}}(\omega)\im\chi_B^{\vphantom{q}}(\omega).
\label{XAB}
\ee
\end{subequations}
Since the divergenceless part\footnote{The
divergenceless part of the Green's dyadic is defined
by $\bm{\Gamma'}=\bm{\Gamma}+\bm{1}$, where $\bm 1$ includes a spatial
delta function.}
 of the vacuum Green's dyadic is
\be
\bm{\Gamma}^{(0)\prime}(\mathbf{r-r'};\omega)=
(\bm{\nabla\nabla}-\bm{1}\nabla^2)\frac{e^{i\omega R}}{4\pi R}
=\left[\mathbf{\hat R\hat R}(3-3i\omega R-\omega^2R^2)-\bm{1}(1-i\omega R-
\omega^2R^2)\right]\frac{e^{i\omega R}}{4\pi R^3},\label{gammaprime}
\ee
where $R=|\mathbf{R}|$ and $\mathbf{\hat R}= \mathbf{R}/R$,
and the matrix trace appearing in Eq.~(\ref{IAB}) is
\begin{subequations}
\be
\im\Gamma_{ji}(-\mathbf{R};\omega)\im\Gamma_{ij}(\mathbf{R};\omega)
=\frac{2\Delta(\omega R)}{(4\pi R^3)^2},\ee
with
\be
\Delta(\omega R)=
(3-2\omega^2R^2+\omega^4R^4)\sin^2\!\omega R-\omega R(3-\omega^2R^2)
\sin 2\omega R+3\omega^2R^2 \cos^2\!\omega R.
\ee
\end{subequations}
Note that this trace is finite as $R\to0$ since
\begin{subequations}
\be
\Delta(u)\sim\frac23 u^6-\frac29 u^8+\cdots, \quad u\ll1.\label{smallu}
\ee
Although it therefore appears  that the leading term, proportional to $u^6$, 
never contributes
to the geometrical integral (\ref{IAB}), this depends on the convergence
of the integrals---see Sec.~\ref{sec:bbmf}.
And, apart from subdominant oscillating terms, the large-argument behavior is
\be
\Delta(u)\sim\frac12u^4+\cdots,\quad u\gg 1.\label{largeu}
\ee
\end{subequations}

An alternative, momentum-space, formulation of the force is presented in
Appendix A.
\section{Inhomogeneous needle}
\label{sec:ihn}
This simplest example is a one-dimensional needle, lying along the $z$-axis,
with one side being made of
uniform material $A$, and of length $a$;
and the other side of uniform  material $B$, and of length $b$: 
\be \chi(\mathbf{r};\omega)=C\left[\delta(x)\delta(y)\theta(z)\theta(a-z)
\chi_A^{\vphantom{q}}
(\omega)+\delta(x)\delta(y)\theta(-z)\theta(b+z)\chi_B^{\vphantom{q}}(\omega)
\right],
\ee
for the needle having constant (small) cross-sectional area  $C$. 
Then, the double volume
integral $I_{AB}$ just involves $R=z'-z$ ($z'>0>z$), and 
can be explicitly evaluated:
\be
I_{AB}=\frac{C^2\omega^5}{(4\pi)^2}\left[f(\omega(a+b))-f(\omega a)
-f(\omega b)\right],
\ee
where,  in terms of a dimensionless variable $x$,
\begin{subequations}
\be
 f(x)\equiv\frac1{30x^5}g(x),\ee with
\be 
g(x)\equiv
-9-5x^2(1+3x^2)+(9-13x^2+11x^4)\cos2x+2x(9-x^2)\sin2x
+22x^5 \Si(2x).
\ee
\end{subequations}
Here, the sine integral function is defined by
\be
\Si(x)=\int_0^x\frac{dt}t \sin t.
\ee
The behavior of the function $g$ for small and large arguments is
\begin{subequations}
\be
g(x)\sim 20x^6-\frac{20}9 x^8 \quad\mbox{as}\quad x\to0, \label{glimits}
\ee
and
\be
 g(x)\sim 11\pi x^5\quad \mbox{as}\quad
x\to\infty.\label{limits}
\ee
\end{subequations}
Thus, the self-propulsive force (\ref{forceAB}) on the needle can be written as
\be
F_z=\frac{C^2}{4\pi^3}\int_0^\infty d\omega\,\omega^5
X_{AB}^{\vphantom{q}}(\omega)
\left[\frac1{e^{\beta\omega}-1}-\frac1{e^{\beta'\omega}-1}\right]
[f(\omega(a+b))-f(\omega a)-f(\omega b)].
\ee

One curious feature is immediately apparent:  For high temperatures, or
$\beta$, $\beta' \ll1$, the frequency integral should be dominated by 
large values of $\omega$, where $f(x)\sim 11/30$  is independent of $x$.
Thus, in this limit, the force is independent of the dimensions of the needle;
that is, only a relatively small region\footnote{At high 
temperatures, because of the thermal factor, high frequencies are most
relevant, and therefore only short distances close to the interface play a
role.  Thus, the size of the object becomes irrelevant.  At low temperatures,
on the other hand, low frequencies are most important, so the force 
is sensitive to the size of the object.}
 of the needle near the intersection
of the two halves contributes to the force.
At room temperature,
this independence of the size of the needle should hold if $a, b\gg 10$ 
$\mu$m.\footnote{By using the conversion factor
$\hbar c=2\times 10^{-5}$ eV\,cm,  $aT\sim 1$ is converted into this value
at room temperature.}
  For low temperatures, or $\beta, \beta'\gg1$, low values of $\omega$
should dominate, and in view of the behavior in Eq.~(\ref{glimits}, the 
leading
term in $g$ will give no contribution to the force, as already recognized
above, and the second term will
yield bilinear behavior, that is $F_z\propto a b$.

Now let us compute the force for a situation with $\chi_A^{\vphantom{q}}$ 
being a real  constant (because the frequencies involved 
should be small compared to the atomic binding energies---but see the
discussion at the end of  Sec.~\ref{sec:jb}),
 while $\chi_B^{\vphantom{q}}$ is given by the Drude model, 
which is a good approximation for a metal:
\be
\chi_B^{\vphantom{q}}(\omega)=-\frac{\omega_p^2}{\omega^2+i\omega\nu},
\label{drude}
\ee
where $\omega_p$ is the plasma frequency  and
$\nu$ is the damping parameter, about 9 eV and 0.035 eV, respectively,
 for gold \cite{Lambrecht}. 
Let us express the dimensional quantities in terms of reference scales,
$\beta_0$ and $a_0$.  For numerical work we will take $\beta_0=40$ (eV)$^{-1}$,
corresponding to room temperature, and a macroscopic size $a_0=1$ cm. For these
values, the dimensionless ratio  $\frac{a_0}{\beta_0}=1250$.
 Then, with 
\be
\beta=\hat\beta \beta_0,\quad \beta'=\hat\beta' \beta_0, \quad \beta_0\omega
=y, \quad\beta_0\nu=\lambda,
\ee
we define
\be
h(u,\hat\beta)=\frac1{u^5}\int_0^\infty \frac{dy}y\frac1{y^2+\lambda^2}
g\left(\frac{a_0}{\beta_0} y u\right)\frac1{e^{\hat\beta y}-1}, 
\ee
in terms of which the force is
\begin{subequations}
\be
F^{\rm IN}=
-\frac{C^2\omega_p^2\nu\chi_A^{\vphantom{q}}}{120\pi^3}
\frac{\beta_0^2}{a_0^5}\hat{F}^{\rm{IN}},\label{needleforce}
\ee
where
\be \hat{F}^{\rm{IN}}=
h\left(\frac{a+b}{a_0},\hat\beta\right)-
h\left(\frac{a}{a_0},\hat\beta\right)-h\left(\frac{b}{a_0},\hat\beta\right)
-h\left(\frac{a+b}{a_0},\hat\beta'\right)
+h\left(\frac{a}{a_0},\hat\beta'\right)
+h\left(\frac{b}{a_0},\hat\beta'\right).\label{needleF}
\ee
\end{subequations}

\begin{figure}
\includegraphics{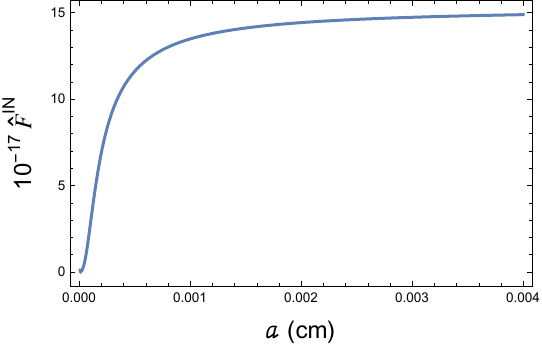}
\caption{Dimensionless force $\hat{F}^{\rm IN}$ 
on an inhomogeneous needle with $a=b$ as a function of $a$ 
for $T=300$ K and 
$T'=600$ K.  The half-length  of the needle is given in cm.
The asymptotic value 
of $\hat{F}^{\rm IN}$ is $1.53\times 10^{18}$.}
\label{fig1}
\end{figure}

\begin{figure}
\includegraphics{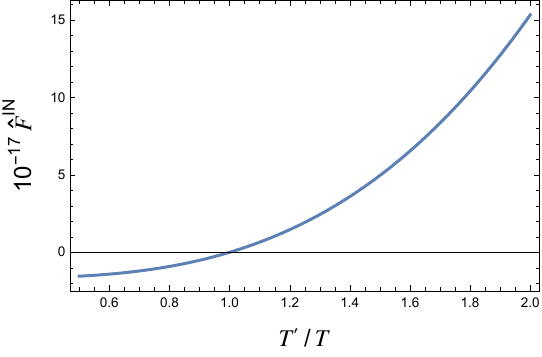}
\caption{Dimensionless force $\hat{F}^{\rm IN}$ 
on an inhomogeneous needle with $a=b$ as a function of $T'$ 
for $T=300$ K.  The half-length  of the needle is taken to be 1  cm.}
\label{fig3}
\end{figure}

We now show numerical results.  We take the above stated parameters
so, for gold, $\lambda=1.4$,
and the magnitude of the prefactor in the force (\ref{needleforce}) 
is 
$1.9\chi_A \times 10^{-20}$\,N
if the cross-sectional radius of the needle is 1 mm.
  Figure \ref{fig1} 
shows the saturation of the force at a value of about 
%0.06
$0.03 \chi^{\vphantom{q}}_A$ N for large $a$.
The figure also shows how the force vanishes for small $a$.
%Figure \ref{fig2} shows the same but for shorter lengths in the sub-micrometer
%range.  The forces are still quite substantial.
%\begin{figure}
%\includegraphics{needlesmalla1.eps}
%\caption{Dimensionless Force $\hat F^{\rm IN}$  on an inhomogeneous
%needle with $a=b$ as a function of $a$ for $T=300$ K and 
%$T'=600$ K.  The length of each half of the needle is given in cm.}
%\label{fig2}
%\end{figure}
Finally, the temperature dependence is sketched in Fig.~\ref{fig3}.

However,  this is likely to be a
gross overestimate. The perturbative estimate for the
susceptibility, given by the Drude model (\ref{drude}),
 would not be expected to
be valid deep within the metal, because of the screening effect of
the other dipoles, so some version of the Lorenz-Lorentz model might
be more appropriate \cite{ce}.  To ameliorate this difficulty, we imagine that 
the metal is thin, so that all points within the body are close to the surface.
The corresponding thickness of the  metal is set by the  skin depth, which we
estimate in terms of the conductivity $\sigma$ as follows \cite{ce},
\be
\delta=(\omega \sigma/2)^{-1/2}=(\omega^2\im\chi/2)^{-1/2}
=\sqrt{\frac{2(\omega^2+\nu^2)}{\omega\omega_p^2\nu}},
\ee
which, for our nominal values and for $\omega\sim \beta_0^{-1}$, turns 
out to be of order 50 nm.  For thicker objects, one would
have to go beyond the perturbative approach adopted here.
So, the force
on a needle with this tiny cross section, rather than one of 1 mm radius,
would be 
reduced by a factor of $\sim10^{-17}$.  Interestingly, 
since $1/\beta_0$ and $\nu$
are comparable, the skin depth is near its minimal value, achieved when
$\omega=\nu$, $\delta_{\rm min}=2/\omega_p$.

The reason the force is toward the metal end of the needle 
(negative in our convention) might seem to be that
emissivity of a metal is far lower than that of an insulator, so there
is more radiant energy emitted from the insulator end of the needle.
The corresponding momentum imbalance drives the object in the $-z$ direction.
However, this is not quite borne out by our perturbative  calculations,
which show, for modest temperatures, more thermal radiation is emitted
by the metal than by the dielectric.  However, the emission from the
dielectric is nearly entirely reflected by the metal, so the momentum
imbalance remains the same.  An explicit example of this is
detailed in Appendix \ref{appc}.

\section{Thin spherical shell}
\label{sec:shell}
In view of the preceding remarks concerning skin depth, we next consider
a spherical shell, with thickness $t$, smaller than the skin depth, and much
smaller than the radius $a$ of the sphere.\footnote{
Hollow gold nanospheres are becoming of great interest for biomedical
applications.  For an example of their synthesis, see Ref.~\cite{synthesis}.}
 We imagine that the sphere is
inhomogeneous: in terms of the polar angle $\theta$,
\be
\chi(\mathbf{r};\omega)=\left\{\begin{array}{cc}
0<\theta<\frac{\pi}2:& \chi_A^{\vphantom{q}}(\omega),\\
\frac{\pi}2<\theta<\pi:&\chi_B^{\vphantom{q}}(\omega).\end{array}\right.
\ee
The square of the distance between two points on the two hemispheres is
given by
\be
R^2=2a^2(1-\cos\gamma)=
2a^2[1-\cos\theta\cos\theta'-\sin\theta\sin\theta'\cos(\phi-\phi')],
\ee
where $\gamma$ is the angle between two points, one on the upper hemisphere
and one on the lower.
Then, the geometrical integral is
\be
I_{AB}=\int_A(d\mathbf{r})\int_B(d\mathbf{r'})\,\partial_z
\left[\frac{\Delta(\omega R)}{(4\pi R^3)^2}\right]
=\frac{\omega^8 a}{(4\pi)^2}
\int_A(d\mathbf{r})\int_B(d\mathbf{r'})\frac{\cos\theta-\cos\theta'}{u} D(u),
\ee
where $u=\omega R$.
The $u$ derivative occurring here is explicitly
\be
D(u)=\frac\partial{\partial u}
\left[\frac{\Delta(u)}{u^6}\right]=
\frac1{u^7}\left[-9-u^2(2+u^2)+(9-16u^2+3u^4)\cos 2u+u(18-8u^2+u^4)
\sin 2u\right],
\ee
which behaves as
\begin{subequations}
\bea
D(u)&\sim& -\frac{4}{9}u,\quad u\to 0,\label{lowu}\\
D(u)&\sim&-\frac1{u^3}+\frac{\sin(2u)}{u^2},\quad u\to \infty.\label{highu}
\eea
\end{subequations}
The low-$u$ behavior follows from Eq.~(\ref{smallu}), where again we note that
 the $u^6$ term does not contribute.
The last term for large $u$ rapidly oscillates, and on the average gives
a term of $O(u^{-4})$, so is subdominant.

\begin{figure}
\centering
\includegraphics{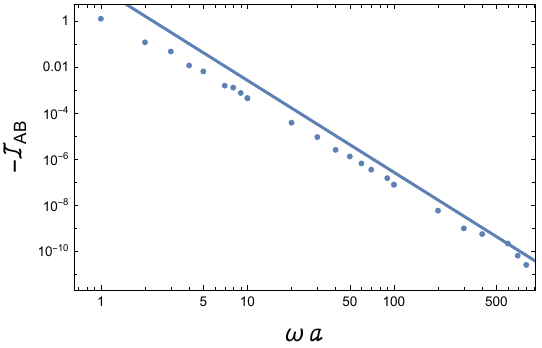}
\caption{The integral $-\mathcal{I}_{AB}$ in Eq.~(\ref{IABsph}), 
 evaluated numerically (blue dots),  The line shows the result
of replacing $D(u)$ by its asymptotic value $-1/u^3$,
which is exactly described by a power law $(\omega a)^{-4}$.   
The coefficient of the fit  to $\mathcal{I}_{AB}$
is  $s=-26.88$. 
The ``exact'' results exhibit considerable scatter, only part of which 
reflects the
serious oscillations of the integrand; the 
numerical integration becomes unstable for
 large values of $\omega a$.  Although those points lie below the
asymptotic line, they seem to be approaching it. }
\label{isph.fig}
\end{figure}

For our inhomogeneous sphere, with vanishing thickness $t$, the geometrical
integral is a threefold one:
\begin{subequations}
\be
I_{AB}=\frac{\omega^8 a^5 t^2}{8\pi}\mathcal{I}_{AB}(\omega a),
\label{IABhs}
\ee
with
\be
\mathcal{I}_{AB}(\omega a)=
\int_0^{\pi/2}d\theta\sin\theta
\int_0^{2\pi} d\phi \int_{\pi/2}^\pi d\theta'\sin\theta'
(\cos\theta-\cos\theta')\frac{D(u)}u.\label{IABsph}
\ee
\end{subequations}
For a macroscopic object, one would think that the large $u$ behavior of $D$
is dominant, because for a 1~cm sphere at room temperature the characteristic
value of $\omega a\sim 1250$.  But this might be misleading, since polar angles
close to the equator of the sphere correspond to $\gamma\to 0$. 
In Fig.~\ref{isph.fig} we show the result of numerical integration, and 
compare with that arising  from the leading large-$u$ behavior.

\begin{figure}
\includegraphics{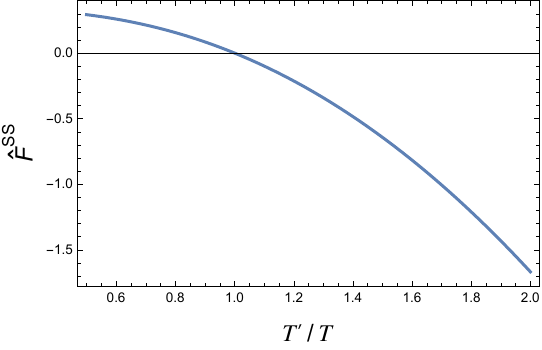}
\caption{Dimensionless force on an inhomogeneous sphere with the northern 
hemisphere
having a constant susceptibility, and the southern hemisphere being a Drude 
metal,
with the environment being at room temperature, for various values of the
temperature of the sphere, $T'$. The prefactor in the force (\ref{forcesph}), 
for our nominal values for gold, and the minimum value of the skin depth, 
$t=2/\omega_p$, is $1.2 \times 10^{-12}\chi_A^{\vphantom{q}}$ N, for a 
sphere of 1 cm radius.
The force is computed in the approximation that the geometric integral $I_{AB}$
is described by the fit to the large-$u$ asymptotics
in Fig.~\ref{isph.fig}.}
\label{fig:stp}
\end{figure}

Using the result of the approximate fit for $\mathcal{I}_{AB}$, 
and the model we used for the 
needle, that $\chi_A^{\vphantom{q}}$ is constant and 
$\chi_B^{\vphantom{q}}$ is given by the Drude model (\ref{drude}), we find
for the self-propulsive force (\ref{forceAB})
\begin{subequations}
\label{forcess}
\be
F^{\rm SS}=
-\frac{\omega_p^2 t^2 a}{16\pi^2}\chi_A^{\vphantom{q}}\nu^3 s
\hat{F}^{\rm SS}, \label{forcesph}
\ee
where $s$ is the slope of the line shown in
Fig.~\ref{isph.fig} and
\be
\hat{F}^{\rm SS}=\int_0^\infty
dx\frac{x^3}{x^2+1}\left(\frac1{e^{\beta\nu x}-1}-\frac1{e^{\beta'x\nu}-1}
\right),\label{hatss}
\ee
\end{subequations}
 the integrals being explicitly
\be
\int_0^\infty dx\frac{x^3}{x^2+1}\frac1{e^{x y}-1}=\frac{\pi^2}{6y^2}-
\frac12\left[\ln\left(\frac{y}{2\pi}\right)
-\frac\pi{y}-\psi\left(\frac{y}{2\pi}\right)\right].
\ee
The dimensionless force $\hat{F}^{\rm SS}$ is plotted in Fig.~\ref{fig:stp}.

\section{Janus Ball}
\label{sec:jb}
We now turn to what has been called a Janus ball, a spherical volume with
the upper and lower halves composed of different homogeneous materials.
In view of our previous remarks about skin depth, we should restrict attention
to a nanometer-sized object.
Because $I_{AB}$ is the integral of a gradient, the double radial integral 
becomes a surface integral,
\be
I_{AB}=\oint_{S_A} d\mathbf{S\cdot \hat z}\int_B(d\mathbf{r'})
\frac{\Delta(\omega R)}{(4\pi R^3)^2},
\ee
where $S_A$ is the surface of the $A$ half-ball, that is, a 
hemisphere of radius
$a$, $0<\theta<\frac\pi2$ and a disk $\theta=\pi/2$, $0\le r\le a$,
while the second integral is 
over the volume of the lower $B$ half-ball, for polar
angle $\frac\pi2>\theta>\pi$.  Explicitly, the integral can be written in two
parts:
\begin{subequations}
\be
I_{AB}=\frac1{8\pi a}\mathcal{I}_{AB},
\ee
with 
\bea
\mathcal{I}_{AB}&=&
\int_0^{\pi/2}d\theta\sin\theta\cos\theta
\int_0^1 d\rho'
\,\rho^{\prime2}\int_{\pi/2}^\pi d\theta'\,\sin\theta'\int_0^{2\pi}d\phi'
\frac{\Delta(\omega a P_1)}{P_1^6}\nn\\
&&\quad\mbox{}-
\int_0^1d\rho\,\rho\int_0^1 d\rho'\,\rho^{\prime2}\int_{\pi/2}^\pi d\theta'\,
\sin\theta'\int_0^{2\pi}d\phi' \frac{\Delta(\omega a P_2)}{P_2^6},
\label{JBIABnum}
\eea
\end{subequations}
where $\rho=r/a$, $\rho'=r'/a$, and
\begin{subequations}
\bea
P_1^2&=&1+\rho^{\prime2}-2\rho'(\cos\theta\cos\theta'+\sin\theta\sin\theta'
\cos\phi'),\\
P_2^2&=&\rho^2+\rho^{\prime2}-2\rho\rho'\sin\theta'
\cos\phi',
\eea
\end{subequations}
because, on the bisecting disk, $\theta=\pi/2$.
Once these four-fold integrals are done, the force, for the same model 
considered above,  is obtained from
\be
F^{\rm JB}=-\frac{\omega_p^2\epsilon}{2\pi^2}\chi_A\int_0^\infty du
\frac1{u(u^2+\epsilon^2)}\mathcal{I}_{AB}(u)\left[\frac1{e^{\beta u/a}-1}
-\frac1{e^{\beta'u/a}-1}\right],\label{forcejs}
\ee
in terms of the abbreviations %\textcolor{red}{$p=\omega_p a$}, 
$\epsilon=\nu a$, $u=\omega a$.

\begin{figure}
\includegraphics{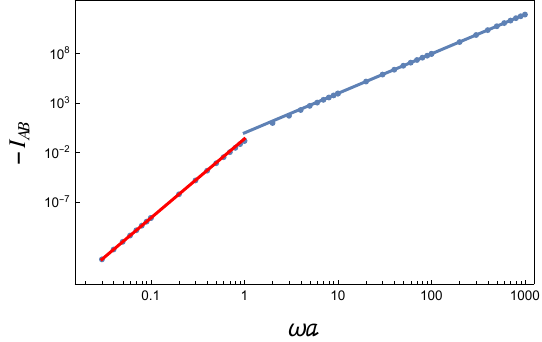}
\caption{The dots show the results of numerical integration of the two
four-fold integrals in (\ref{JBIABnum}). Shown is the magnitude of
the difference of the two integrals.  The upper line (blue) is the large-$u$ 
asymptotic behavior resulting from (\ref{largeu}), 
that is, $\mathcal{I}_{AB}\sim -0.927 (\omega a)^4$. The fit is very good
for  values of $\omega a>10$.  As seen from the lower
line (red), values of $\omega a$ smaller  than
1 are well described by $\mathcal{I}_{AB}\sim -\frac{2\pi}{27}(\omega a)^8$,
which results from the $u^8$ term in Eq.~(\ref{smallu}), 
the $u^6$ term never contributing.}
\label{JBiab}
\end{figure}

Numerical integration of the two integrals in Eq.~(\ref{JBIABnum}) is shown in 
Fig.~\ref{JBiab}.
The figure shows that for large values of $\omega a$, appropriate to a 
macroscopic ball near room temperature, the power-law 
fit from the large 
$u$ behavior of $\Delta(u)$ in Eq.~(\ref{largeu}) is quite good, and results
in the following behavior of $I_{AB}$:
\be
I_{AB}^{\rm JB}\sim-\frac1{8\pi a}(0.927)(\omega a)^4.
\ee
This has the same $\omega$ dependence as that for a thin spherical shell in
this limit, which, from Eq.~(\ref{IABhs}), is
\be
I_{AB}^{\rm HS}\sim -\frac{26.9}{8\pi}\omega^4 a t^2.
\ee
This will be the same as $I_{AB}^{\rm JB}$ if $a=5.39t$.  So apart from
a factor of $0.035 (a/t)^2$, the forces will be the
 same in the large $a$ limit.  

\begin{figure}
\includegraphics{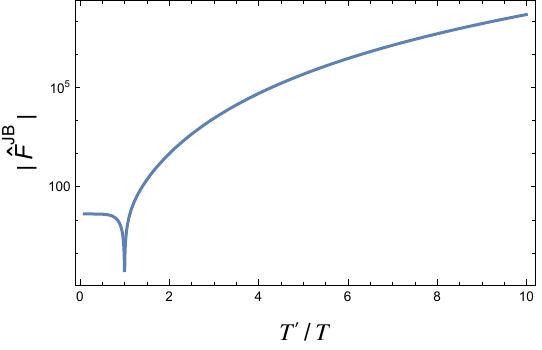}
\caption{Dimensionless force $\hat{F}^{\rm JB}$ (\ref{JBForcelow}) 
on a Janus ball
composed of an upper hemisphere made of a dispersionless dielectric, and
a lower hemisphere composed of a Drude metal.  Parameters are as given
in the text.  The force, given by Eq.~(\ref{JBForcelow}), is plotted  as a
function of the temperature of the sphere, $T'$, for the temperature of
the environment fixed at room temperature.  The magnitude is shown in this
log plot; the force is negative for $T'>T$.}
\label{figFJBlowa}
\end{figure}

\begin{figure}
\includegraphics{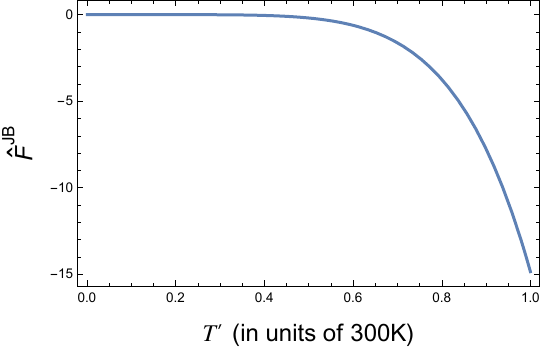}
\caption{Dimensionless force $\hat{F}^{\rm JB}$ (\ref{JBForcelow}) 
for a Janus ball composed of gold and
SiO$_2$, with the vacuum at zero temperature, as a function of the temperature
of the ball expressed in units of room temperature.}
\label{JBF}
\end{figure}

But, as noted above, only the small sphere situation is likely to be
physically relevant.
For small $a$, the forces will have a different behavior.
The lower (red) line in Fig.~\ref{JBiab} is $\mathcal{I}_{AB}=-\frac{2\pi}{27}
(\omega a)^8$, which follows from performing 
the elementary integrations occurring when
Eq.~(\ref{smallu}) holds:  the $u^6$ term cancels between
the disk and the hemisphere surface integrals (as noted above, the $u^6$ term
never contributes), and the $u^8$ term gives the
value stated.  So for a small Janus ball the force is given by
\begin{subequations}
\label{jbforcesmall}
\be
F^{\rm JB}=\frac1{27\pi}\chi_A\omega_p^2(\nu a)^7 
\hat{F}^{\rm JB},\label{JBpre}
\ee
where the dimensionless force is 
\bea
\hat{F}^{\rm JB}&=&
\int_0^\infty dx\frac{x^7}{x^2+1}\left(\frac1{e^{\beta\nu x}-1}-
\frac1{e^{\beta'\nu x}-1}\right)\nn\\
&=&-\frac12\left[\ln\frac{T'}T+\frac\pi\nu(T'-T)+\psi\left(\frac\nu{2\pi T'}
\right)-\psi\left(\frac\nu{2\pi T}\right)\right]\nn\\
&&\quad\mbox{}+\frac{\pi^2}6\frac1{\nu^2}(T^2-T^{\prime2})
-\frac{\pi^4}{15}\frac1{\nu^4}(T^4-T^{\prime4})
+\frac{8\pi^6}{63}\frac1{\nu^6}(T^6-T^{\prime6}).\label{JBForcelow}.
\eea
\end{subequations}
The absolute value of this  function is plotted in Fig.~\ref{figFJBlowa}, 
as a function of the body temperature $T'$, for
the environment temperature $T=300$ K, and $\nu=0.035$ eV, appropriate for
gold.  If, further. the radius of the  sphere is 1 $\mu$m, the prefactor 
for a gold microsphere ($\omega_p=9$ eV) is $3.84 \times 10^{-18} \chi_A$ N, 
so this might be observable.\footnote{The corresponding
acceleration would be of order $10^{-2}$ cm/s$^2$.}
  Note that the force is negative for $T'>T$, so 
it is in the direction of  the metal end of the ball.
These results are roughly consistent with those found in Ref.~\cite{Reid}.
Those authors
 consider a Janus ball with one half SiO$_2$ and the other half gold,
and for $T=0$ K and $T'=300$ K find a force of about $4\times 10^{-18}$ N on a
1 $\mu$m ball.  Here, for the same example, we find the prefactor in 
Eq.~(\ref{JBpre}) to be $3 \times10^{-18}$ N, while $\hat{F}_{\rm JB}$ 
is about $-15$.
A plot of $\hat{F}$ for a vacuum  environment at 0 K is shown in 
Fig.~\ref{JBF}, again qualitatively similar to the results of Ref.~\cite{Reid}.
However, the scaling with the radius of the ball is quite different: we
find a $a^7$ behavior, whereas Ref.~\cite{Reid} reports scaling roughly with the volume.

\begin{figure}
\includegraphics{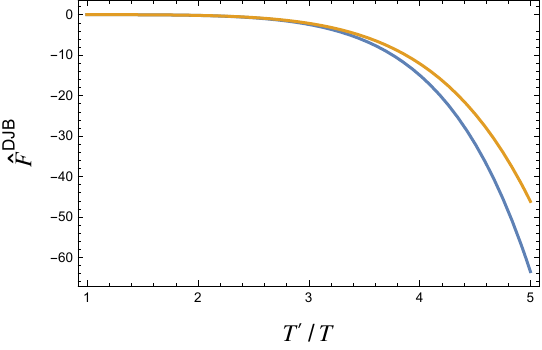}
\caption{Plot of the dimensionless force $\hat{F}^{\rm DJB}$ 
for a dispersive Janus ball with
dielectric on the $A$ side and metal on the $B$ side, as
a function of the temperature of the ball.  The metal is chosen to
be gold, for which $\lambda=1.4$, the dielectric is polystyrene, for which,
roughly \cite{Parsegian} (taking the first term in a four-term fit to the
susceptibility) $y_0=240$ and $\mu=26$, and we consider a room-temperature
background.  The upper (ochre) curve assumes a dispersionless dielectric,
while the lower (blue)
curve includes the single Lorentzian dispersion model.  The 
effect of the latter is rather small
because the relevant frequencies are small
compared to $\omega_0$, except at very high temperature.}
\label{fig:realchi}
\end{figure}

For sufficiently high temperatures, the dispersion of the
dielectric cannot be ignored.  Therefore, let us
 use a more realistic Lorentzian model,
\be
\chi_A^{\vphantom{q}}(\omega)=\frac{\tilde{\omega}_p^2}{\omega_0^2-\omega^2-i
\omega\gamma}.
\ee
Then, the second-order susceptiblity factor 
(\ref{XAB}) that appears in the force is
\be
X_{AB}(\omega)=-\frac{\omega_p^2\tilde{\omega}_p^2\left[\omega_0^2\nu
+\omega^2(\gamma
-\nu)\right]}{\left[(\omega_0^2-\omega)^2+\omega^2\gamma^2\right]\left[\omega
(\omega^2+\nu^2)\right]}.
\ee
Of course, this agrees with what we had previously if $\omega\ll\omega_0$,
provided $\chi_A^{\vphantom q}=\tilde{\omega}_p^2/\omega_0^2$.
The propulsive force is now for a small dispersive
 Janus ball, where $I_{AB}=-\frac{(\omega a)^8}{108 a}$:
\begin{subequations}
\be
F^{\rm DJB}=\frac1{27\pi}\omega_p^2\tilde{\omega}_p^2 a^7T_0^5
\hat{F}^{\rm DJB},
\ee
 the dimensionless force being 
\be
\hat{F}^{\rm DJB}=\int_0^\infty dy\, y^7\frac{y_0^2\lambda+y^2(\mu-\lambda)}
{[(y_0^2-y^2)^2+y^2\mu^2][y^2+\lambda^2]}\left[\frac1{e^{y/t}-1}
-\frac1{e^{y/t'}-1}\right].
\ee
Here, in terms of the inverse of room temperature, $\beta_0$,
the dimensionless variables are
\be
\lambda=\beta_0 \nu,\quad \mu=\beta_0\gamma, \quad y_0=\beta_0\omega_0,
\quad y=\beta_0 \omega, \quad t=\beta_0 T,\quad t'=\beta_0 T'.
\ee
\end{subequations}
We illustrate the effect of this change on the force in Fig.~\ref{fig:realchi}.

\section{Blackbody/Metal Planar Surface}
\label{sec:bbsurf}
\subsection{Blackbody surface}
We preface this section by considering the power absorbed by an object out
of thermal equilibrium.  Classically, the power is
\be
P=\frac{\partial \mathcal{F}}{\partial t}
=-\int (d\mathbf{r})\mathbf{P}(\mathbf{r},t)
\cdot\frac\partial{\partial t}\mathbf{E}(\mathbf{r},t).
\ee  
We see here the reappearance of the (free) energy density 
(\ref{freeen}), $\mathcal{F}=\int (d\mathbf{r}) \mathfrak{f}$.
For an object at temperature $T'$ in vacuum at temperature $T$, the
FDT implies, in first order,
\be
P=\int (d\mathbf{r})\frac{d\omega}{2\pi}\omega \im \chi_{ij}(\mathbf{r};\omega)
\im \Gamma_{ji}(\mathbf{r,r'};\omega)\bigg|_{\mathbf{r'\to r}}\left[\coth
\frac{\beta\omega}2-\coth\frac{\beta'\omega}2\right].\
\ee
For an isotropic body, $\chi_{ij}(\mathbf{r};\omega)=\chi(\mathbf{r};\omega)
\delta_{ij}$, and a vacuum environment, where $\tr\im\mathbf{\Gamma}
(\mathbf{r,r'}; \omega)\big|_{\mathbf{r=r'}}=\omega^3/(2\pi)$, 
the power simplifies to
\be P=\frac1{\pi^2}\int(d\mathbf{r})\int_0^\infty d\omega\,\omega^4
\im\chi(\mathbf{r};\omega)\left[\frac1{e^{\beta\omega}-1}
-\frac1{e^{\beta'\omega}-1}\right].\label{power}
\ee

Now the power absorbed by a blackbody is given by Stefan's law,
\be
P=S \frac{\pi^2}{60}(T^4-T^{\prime 4}),
\ee
 where $S$ is the surface area of
the blackbody.  Since the radiation from a blackbody arises entirely from
the surface, we can write
\be
\int_V (d\mathbf{r})\chi(\mathbf{r};\omega)=\int_S dS \tilde \chi(\mathbf{r};
\omega),
\ee
so in order for Eq.~(\ref{power}) to describe a blackbody, the surface
susceptibility $\tilde{\chi}$ must satisfy 
\be
\im \tilde\chi(\omega)=\frac1{4\omega} \Rightarrow \tilde \chi(\omega)
=\frac{i}4\frac1{\omega+i\epsilon}=\frac\pi4\delta(\omega)
+\mathcal{P}\frac{i}{4\omega},
\ee
$\mathcal{P}$ signifying the Cauchy principal value, 
the latter following from the Kramers-Kronig relation.

\subsection{Force on a blackbody/metal plate}
\label{sec:bbmf}
\begin{figure}
\includegraphics{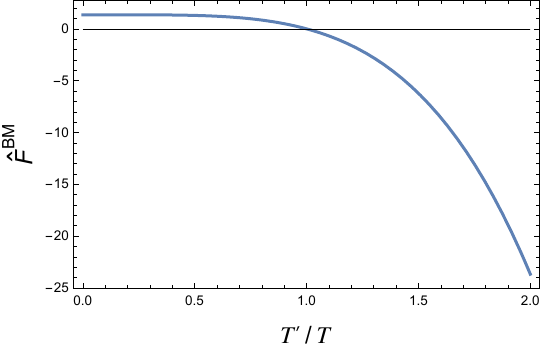}
\caption{Dimensionless force (\ref{JBF2})
on a thin plate consisting of a blackbody on
the positive-$z$ side and a Drude metal on the negative-$z$ side.
The background vacuum is at room temperature.}
\label{fig:FJP}
\end{figure}

Now we want to calculate the self-propulsive force on a composite plate 
consisting
of a blackbody surface on one side and a Drude metal on the other.  The force
is given by Eq.~(\ref{forceAB}), in terms of the geometrical integral
(\ref{IAB}),
\be I_{AB}=
\int_A(d\mathbf{r})\int_B(d\mathbf{r'})\nabla_z
\frac{\Delta(\omega|\mathbf{r-r'}|)}{(4\pi|\mathbf{r-r'}|^3)^2}.
\ee
Using polar coordinates for the transverse spatial integrals, and integrating
out the $z$ derivative, we find for plates of (large) cross sectional area $S$
and thicknesses of the two sides $t_A$ and $t_B$, respectively,
\bea
I_{AB}&=&\frac{S}{8\pi}
\int_0^\infty \rho\,d\rho \int_{-t_B}^0 dz'
\left[\frac{\Delta\left(\omega\sqrt{\rho^2+(t_A-z')^2}\right)}
{(\rho^2+(t_A-z')^2)^3}-
\frac{\Delta(\omega\sqrt{\rho^2+z^{\prime2}})}{(\rho^2+z^{\prime2})^3}\right]
\nn\\
&=&\frac{S}{8\pi}\int_{-t_B}^0 dz'\int_{t_A-z'}^{-z'}du\frac{\Delta(\omega u)}
{u^5}.
\eea
Because $\omega\sim T,\,T'$ corresponds to a length scale
$\sim 10\,  \mu$m for temperatures around room
temperature, for thicknesses much smaller
than that, we can expand $\Delta(\omega u)$ using only the first term in
Eq.~(\ref{smallu}), and then we obtain immediately\footnote{
For this planar geometry, it might seem more  convenient to use 
a ($2+1$)-dimensional breakup,
\be
\Gamma_{ij}(\mathbf{r,r'};\omega)=\int\frac{(d\mathbf{k}_\perp)}{(2\pi)^2}
e^{i\mathbf{k}_\perp\cdot(\mathbf{r-r'})_\perp}
g_{ij}(z,z';\mathbf{k}_\perp,\omega).
\ee
Carrying out the trace over the Green's functions, and
simplifying to the case that the 
susceptibility is independent of $x$ and $y$, we readily obtain
\be
I_{AB}=-\frac{S}{4}\int_{|\mathbf{k}_\perp|<|\omega|}
\frac{(d \mathbf{k}_\perp)}{(2\pi)^2}
\int_A dz\int_B dz' \frac{\lambda^4+k^4}{\lambda}\sin2\lambda(z-z')
\ee
Now, if the two sides of the thin plate have small thickness $t_A$ and $t_B$, 
respectively, we can expand the sine for small $z$ and $z'$ and find
the same result as Eq.~(\ref{IABtp}).}
\be
I_{AB}=-\frac{S}{24\pi}t_At_B(t_A+t_B)\omega^6.\label{IABtp}
\ee
Then the force is given by
\be
F_z=-\frac{S}{6\pi^2} t_A t_B(t_A+t_B)\int_0^\infty d\omega
X_{AB}(\omega)\omega^6\left[\frac1{e^{\beta\omega}-1}
-\frac1{e^{\beta'\omega}-1}\right].
\ee
When $A$ is the blackbody, $t_A\im\chi_A^{\phantom{q}}=\frac1{4\omega}$, while
$B$ is the Drude metal, described by Eq.~(\ref{drude}),
the force is
\be
F_z=\frac{S t_B(t_A+t_B)}{24\pi^2}\omega_p^2\int_0^\infty d\omega
\frac{\omega^5}{\omega^2+\nu^2}\left[\frac1{e^{\beta\omega}-1}
-\frac1{e^{\beta'\omega}-1}\right].
\ee
Again, this is of the same class of functions explored in Appendix A of
Ref.~\cite{QVT},
\begin{subequations}
\be
F_z^{\rm BM}=\frac{S t_B(t_A+t_B)\omega_p^2 \nu^4}{24\pi^2}\hat{F}^{\rm BM},
\label{JPF1}
\ee
where, in terms of the digamma function, 
$\psi(z)=\Gamma'(z)/\Gamma(z)$,
\bea
\hat{F}^{\rm BM}&=&
\int_0^\infty dx\frac{x^5}{x^2+1}\left[\frac{1}{e^{\beta\nu x}-1}
-\frac1{e^{\beta'\nu x}-1}\right]\nn\\
&=&\frac{\pi}{2\nu}(T'-T)+\frac12\ln\frac{T'}T+\frac12\psi\left(\frac\nu{2\pi
T'}\right)-\frac12\psi\left(\frac\nu{2\pi T}\right)-\frac{\pi^2}{6\nu^2}(T^2-
T^{\prime 2})+\frac{\pi^4}{15\nu^4}(T^4-T^{\prime 4}).\label{JBF2}
\eea
\end{subequations}
This is plotted in Fig.~\ref{fig:FJP}.
The values there are
 to be multiplied by the prefactor in Eq.~(\ref{JPF1}), which for the
area of the plate being $S=1$~cm$^2$, $t_A=t_B=10$ nm, 
and, appropriate for gold,
$\omega_p=9$ eV, $\nu=0.035$ eV, is 
 $\sim 10^{-13}$ N.  Again, this force should
be observable, {since it corresponds to an acceleration of 
$\sim40$ m/s$^2$.
The results in this section are comparable to those found
recently in Ref.~\cite{manjavacas}, which concerns the force on a similar
two-layer planar structure.
In particular, if we use $t_B=2/\omega_p$, $t_A=0$, and consider
high temperatures (so the last term in Eq.~(\ref{JBF2}) dominates), we recover
the ideal result (3) in that reference.

\section{Friction}
\label{sec:fric}
Of course, once a body is put in motion by the self-propulsive force, it will
experience quantum friction in the vacuum.  This 
friction was considered, in first order, for example,
in Ref.~\cite{guo2}, but primarily in the condition of nonequilibrium
steady state (NESS).  The general formula for the
frictional force on a particle in its rest frame is given by Eq.~(5.7) 
{in Ref.~\cite{guo2},
and shows the effects of field and dipole fluctuations,
\be
F_f^{\rm P}=\frac1{3\pi^2\gamma v}\int_0^\infty d\omega\,\omega^4
\im \alpha_{\rm P}\int_{\gamma(1-v)}^{\gamma(1+v)}dy (y-\gamma)f^{\rm P}(y)
\left(\frac1{e^{\beta \omega y}-1}-\frac1{e^{\beta'\omega}-1}\right),
\label{1stfric}
\ee
where $\rm P$ designates the polarization state of the particle, with the
structure functions being [$\gamma=(1-v^2)^{-1/2}$]
\be
f^{\rm P}(y)=\left\{\begin{array}{cc}
\frac3{2\gamma v},& \mbox{ISO},\\
\frac3{4\gamma v}\left(1-\frac{(y-\gamma)^2}{\gamma^2v^2}\right),&\mbox{ZZ}.
\end{array}\right.
\ee
Here, the particle is described by a mean polarizability,
\be
\bm{\alpha}(\omega)=\int(d\mathbf{r}) \bm{\chi}(\mathbf{r};\omega).
\ee
 ISO means a particle which is isotropically polarizable, and ZZ
means one which is only polarizable in the $z$ direction, the direction
of motion. Because of  the symmetry of the $y$ integrand about $y=\gamma$,
the second term here, due to dipole fluctuations, vanishes;  hence,
there is no dependence on the temperature of the particle.  (See
Appendix \ref{appa} for the second-order correction to the frictional force.)

{We are here considering a body made of isotropic material.
% For a needle moving in the direction of its axis, the ZZ polarization
%would seem to be  appropriate.\footnote{Actually, in this weak-susceptibility
%approximation the susceptibility is assumed isotropic, so it might be more
%accurate to multiply the friction by a factor of 5.
At low velocities, the frictional force for this case reduces to
 the Einstein-Hopf formula \cite{EinsteinHopf}
\be
F_f=-\frac{v\beta}{12\pi^2}\int_0^\infty d\omega\,\omega^5
\im\alpha(\omega)\frac1{\sinh^2\beta\omega/2}.
\ee
%which is just $1/5$ the Einstein-Hopf formula for an isotropic particle.
The essential point here is that $F_f$ is proportional to the velocity in
the low-velocity limit, {that is, $F_f=v F'_f$.

Due to the self-propulsive force, the body, out of equilibrium with its
environment, will begin to accelerate.  It will reach a terminal velocity
$v^{\vphantom{q}}_T$
 when the propulsive force exactly balances the frictional force,
\be
F+v^{\vphantom{q}}_TF'_f=0,\quad v^{\vphantom{q}}_T=-\frac{F}{F'_f}.
\ee
Using the same model for the needle as in Sec.~\ref{sec:ihn}, the derivative
of the frictional force is
\be
F'_f=-\frac{2 Ca\omega_p^2\nu}{3\pi^2\beta^2}\int_0^\infty dx\frac{x^4}{x^2
+(\beta\nu/2)^2}\frac1{\sinh^2x},\label{fpsf}
\ee
where for $\beta=40$ (eV)$^{-1}$ ($T=$ room temperature) and a gold needle,
$\beta\nu/2=0.7$, and we find the $x$ integral here evaluates to 0.90.
Then, from the propulsive force (\ref{needleforce}), we obtain the
terminal velocity 
\be
v^{\vphantom{q}}_T=-\frac{F^{\rm IN}}{F'_f}
=\frac{\chi_A^{\vphantom{q}}\beta_0^2\beta^2}{40\pi}
\frac{C}{a_0a^5}
\frac{\hat{F}^{\rm IN}}{0.90}.
\ee
(The velocity is in the $-z$ direction because $\hat F<0$
if $T'>T$.)
For a needle of 1 cm half-length and radius 10 nm, at a temperature
twice that of the 300 K background,   
this equals $1.5\times 10^{-8}$ or about 4 m/s.
This would seem to be very observable, if the nonequilibrium temperature 
imbalance could be maintained.  

However, it would take an extraordinarily long time to reach this terminal
velocity.  From Newton's law,
\be
v(t)=v^{\vphantom{q}}_T\left(1-e^{-t/t_0}\right),
\ee
where the characteristic time is estimated from above to be
\be
t_0= \frac{m}{F_f'}=
\frac{3\pi^2\beta^2  \rho}{2\omega_p^2\nu}\frac1{0.9},
\ee
 $\rho$ being the density of the metal part of the needle.  Putting
in numbers appropriate for gold,  we find 
$t_0\sim 5\times 10^{8}$ s, about 15
years.  This long time is a
reflection of the weakness of the frictional force.

\section{Relaxation to thermal equilibrium}
\label{sec:thermal}
Quantum friction appears then to play a negligible role.  However, unless
the body is equipped with a mechanism to maintain its elevated temperature,
like a proposed quantum thruster \cite{thruster},
 it  will rapidly cool to the temperature of the
environment.  We use an adiabatic approximation to
estimate this effect as in Ref.~\cite{QVT}: The rate of heat
loss
\be
\frac{dQ}{dt}=C_V(T')\frac{dT'}{dt}=P(T',T),
\ee
where $C_V(T')$ is the specific heat of the body at temperature $T'$, and the
power for a slowly moving body is given by Eq.~(\ref{power}) \cite{guo2},
or its generalization (see Appendix \ref{appb})
\be
P(T',T)= \frac1{3\pi^2}\int_0^\infty d\omega\,\omega^4\im
\tr\bm{\alpha}(\omega)\left[
\frac1{e^{\beta\omega}-1}-\frac1{e^{\beta'\omega}-1}\right],
\ee
 but this transcends the weak susceptibility limit, in that
the polarizability is not  simply
 $\alpha(\omega)=\int(d\mathbf{r})\chi(\mathbf{r},\omega)$.
Then, the time to cool the body from temperature $T'_0$ to temperature $T'_1$ 
is
\be
t_1=\int_{T_0'}^{T'_1} dT'\frac{C_V(T')}{P(T',T)}.
\ee
We can solve this equation (numerically) for $T'_1(t_1,T)$ (assuming fixed
environmental temperature), and then find the body's acceleration in terms
of the propulsive force at that time ($m$ is the mass of the body),\footnote{
In this regime,
Newton's law is certainly valid for a macroscopic or mesoscopic body.}
\be
m \frac{d v(t)}{dt}=F(T'(t),T), 
\ee
which, when integrated over all time, gives the terminal velocity,
\be
v^{\vphantom{q}}_T=\frac1m \int_0^\infty dt\, F(T'(t), T).\label{tervelcool}
\ee
This is precisely the scheme advocated in Ref.~\cite{Reid}.

Let us specifically consider the example of the Janus ball discussed in 
Sec.~\ref{sec:jb}.  For the cooling, we consider the high-temperature 
approximation of the Debye model treated in Ref.~\cite{QVT}, 
together with the Lorenz-Lorentz model 
for the polarizability,  which gives
\be
t_1=t_c \int_{T'_0/T}^{T_1'/T}du\frac1{1-u^6}, \quad
t_c=\frac{21}{8\pi^4}\frac{n \omega_p^2}{\nu T^5},\label{coolt1}
\ee
$n$ being the number density.  The time scale $t_c$ for gold at room
temperature is about 2,000 s. 
The numerical inverse function $T'(t)$ is shown in Fig.~\ref{fig-inv}.
\begin{figure}
\includegraphics{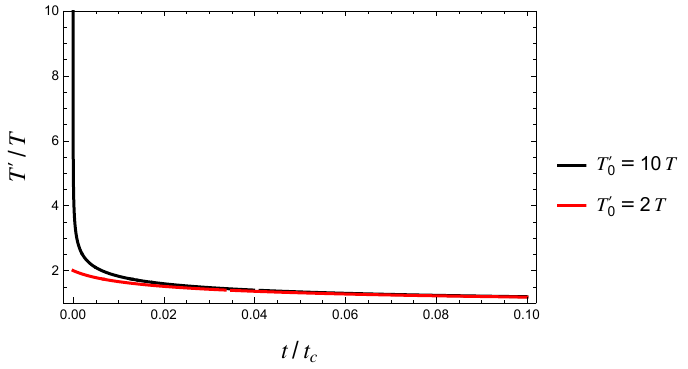}
\caption{The temperature $T'$ of the body (relative to that of the environment
$T$) in terms of the cooling time relative to the scale $t_c$.  The two
curves are for an initial temperature ratio of 10 (upper) and 2 (lower).}
\label{fig-inv}
\end{figure}
Now we insert this into the force formula for the propulsive force on
a small Janus ball, given by Eq.~(\ref{jbforcesmall}).
Then we use the expression for the terminal velocity (\ref{tervelcool}) to give
\be
v^{\vphantom{q}}_T=\lim_{t\to\infty} v(t),\quad v(t)=\frac7{72\pi^5}
\chi_A^{\vphantom{q}}\omega_p^4 \frac{n}m \frac{\nu^6 a^7}{T^5}
\int_0^{t/t_c}d\tau \hat{F}^{\rm JB}(\tau),\quad \tau=\frac{t'}{t_c}.
\ee
The prefactor here, for the metallic half-ball being gold, 
having a radius 100 nm, in a vacuum  environment
at room temperature, is about $20\chi_A^{\vphantom{q}}$ $\mu$m/s. 
The dimensionless force, $\hat{F}^{\rm JB}(t/t_c)$ 
is shown in Fig.~\ref{fig:effhat}.
\begin{figure}
\includegraphics{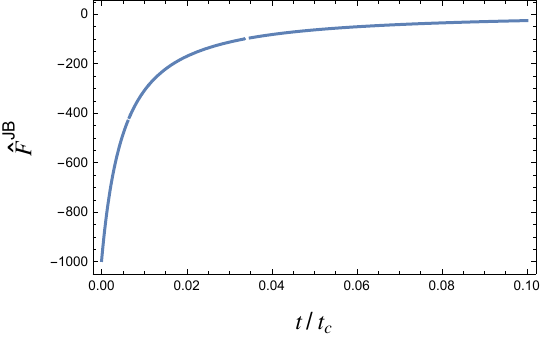}
\caption{The dimensionless force $\hat{F}^{\rm JB}$ for a Janus ball, the metallic half
of which is gold,  as a function
of the time $t/t_c$, at a vacuum temperature of $T=300$ K, and an initial
temperature of the nanoball of $T'_0=600$ K.}
\label{fig:effhat}
\end{figure}
The numerical integral of $\hat{F}^{\rm JB}$ for large $t/t_c$ when $T_0'/T=2$
gives about $-15.6$, yielding roughly  $v^{\vphantom{q}}_T=-300$ 
$\mu$m/s,\footnote{Inversion of $t(T')$ is not necessary
to calculate the terminal velocity.  If we write Eq.~(\ref{coolt1}) for fixed
$T$ as $t_1/t_c = \int_{u_0}^{u_1} du f(u)$ then the terminal velocity is
\be 
v^{\vphantom{q}}_T=\frac{t_c}m \int_{u_0}^1 du\,f(u) F(u).
\ee}
the asymptotic value being achieved by 
about $0.1\, t_c\approx 200$ s.  If there had been no thermal cooling, 
in 2000 s, when the force has dropped nearly to zero, 
the velocity would have been nearly 1800 $\mu$m/s.
%this time  the body would have reached a velocity of about 0.1 cm/s.
Increasing the initial temperature of the particle to $T_0'=10 T$ 
increases the terminal velocity by  a factor of 
about 9, while reducing the
time to reach equilibrium by  a factor of about 20.
(The discussion in this paragraph assumes that the thermal
radiation from the metal portion dominates the energy loss mechanism, which
is true, at least perturbatively, because only the 
susceptibility of the metal has a significant imaginary part}.

\subsection{Weak-susceptibility model}
The above treatment might seem inconsistent, since the 
perturbative susceptibility is used
for the propulsion but the dressed Lorenz-Lorentz polarizability is used for
the cooling.  If, instead, we use the weak-coupling susceptility for the metal,
\be
\im\chi(\omega)=\frac{\omega_p^2 \nu}{\omega(\omega^2+\nu^2)},
\ee
we obtain instead of Eq.~(\ref{coolt1}) the time to cool from $T'_0$ to $T'_1$
\be
t_1=-\frac{6\pi^2 n}{\nu^2\omega_p^2}\int_{\beta'_0\nu}^{\beta'_1\nu}
\frac{dx'}{x^{\prime 2}}\left[\ln\frac{x'}x+\psi\left(\frac{x}{2\pi}\right)-
\psi\left(\frac{x'}{2\pi}\right)+\pi\left(\frac1x-\frac1{x'}\right)+
\frac{\pi^2}3\left(\frac1{x^2}-\frac1{x^{\prime2}}\right)\right]^{-1}.
\label{coolt2}
\ee
The prefactor here, $t_c$,  for gold is about $2\times 10^{-4}$ s.
This is then inverted to find $T'/T$ as a function of 
$\tau=t'/t_c$ as above.
Then the terminal velocity for a Janus ball, half of which is gold,
is found from Eq.~(\ref{tervelcool}) to be
\be
v^{\vphantom{q}}_T=\frac13\frac{\chi_A^{\vphantom{q}}\nu^5 a^4}
{m_{\rm Au}}\int_0^\infty d\tau \hat{F}^{\rm JB}(T'(\tau),T), 
\ee
$m_{\rm Au}$ being the mass of a gold atom, 
where the prefactor is only $\sim 1 $ pm/s, 
again for a radius of 100 nm, while the integral is
only about 100, so this terminal velocity   
would be difficult to observe.\footnote{
Note that the ratio of the $t_c$ scales in Eqs.~(\ref{coolt1}) and 
(\ref{coolt2}) is $\frac7{16\pi^6}\frac{\omega_p^4\nu}{T^5}\sim 10^7$, which
is also the ratio of the forces, apart from the dimensionless integrals,
which are in the ratio of 16/100.}

One might think one could boost this value
substantially by considering a macroscopic
object.  But, presumably it should be thin so that thicknesses do not exceed
the skin  depth, otherwise the weak susceptibility approach to the force will
not be valid.  So let's consider the thin spherical shell model of 
Sec.~\ref{sec:shell} for which the propulsive force is given, approximately, by
Eq.~(\ref{forcess}).  Since the cooling time is independent of the
shape of the body, we can use Eq.~(\ref{coolt2}), and then find the
terminal velocity from 
\be
v^{\vphantom{q}}_T=\frac3{16\pi}\frac{|s|\nu}{m_{\rm Au}}\frac{t}a 
\chi_A^{\vphantom{q}}\int_0^\infty d\tau |\hat{F}^{\rm SS}(T'(\tau),T)|.
\ee
{For  shell thickness $t=2/\omega_p\approx 50$ nm
and  shell radius  $a=1$ cm,
the prefactor evaluates to $4 \times 10^{-10} \chi_A^{\vphantom{q}}$ m/s, 
where the integral over Eq.~(\ref{hatss}) is about 0.4, so this 
terminal velocity 
would again be difficult to measure.

\section{Conclusions}
In this paper we have tried to give a systematic treatment of self-propulsion
of an inhomogeneous body in vacuum, which occurs not in first order, but in
second order in the susceptibility (or polarizability) of the body.  For
a force to occur, the body must have  nonuniform
electric susceptibility, be dissipative, and
be out of thermal equilibrium with its environment.  
For a body hotter than the vacuum environment and having two
different halves, the force is in the direction of the side having the
lowest emissivity, the radiation emitted being predominantly from the opposite
side. We consider four different configurations: a thin needle, a thin
 spherical shell,   a solid nanoball, and finally a thin plate.
Except in the last case, one side is regarded as non-dissipative, while 
the other is modeled as
a Drude metal; for the thin plate, the nonmetal side
 is modelled as a total blackbody absorber.
  Presumably, in all cases, the objects must have small
thicknesses, so that the skin depth is larger than the thickness.\footnote{
This may also validate the weak-susceptibility approximation, in that we
take the polarizability to be
 simply the volume integral of the susceptibility.
For objects thicker than the skin depth, nonpertubative
effects would be expected to come into play, such as the Lorenz-Lorentz
correction, resulting in the reduction of the local electric field seen by
an atom \cite{ce}.}
Therefore, the forces must be very small (not sufficient for a quantum thruster
\cite{thruster}), 
but probably amenable to observation if the temperature
imbalance could be maintained,
which would require some sort of internal or external
influence.  Quantum vacuum friction will eventually
 cause the particle to acquire a terminal velocity.
But if the system is allowed to relax to thermal equilibrium,
the body might reach an observable velocity in a matter of  seconds;
however, this conclusion sensitively depends on the model of 
susceptibility employed, as we have seen.

\begin{acknowledgments}This work was supported in part by the US National 
Science Foundation, grant number 2008417.
GK thanks the Homer L. Dodge Department
of Physics and Astronomy at 
the University of Oklahoma  for its hospitality. We thank 
Xin Guo, Prachi Parashar, and Steve Fulling for
collaborative assistance.  
We thank Alejandro Manjavacas for bringing 
Ref.~\cite{manjavacas}
to our attention and enlightening correspondence. This paper
reflects solely the authors' personal opinions and does not represent
the opinions of the authors' employers, present and past, in any
way. For the purpose of open access, the authors have applied a
CCBY public copyright license to any Author Accepted Manuscript version
arising from this submission.
\end{acknowledgments}

\appendix
\section{Force on moving body}
\label{appa}
In this appendix we derive the force on a body moving with velocity
$\mathbf{v}$ in the $z$ direction, which in general will include both a 
propulsive force and a frictional force.  For simplicity of presentation,
we will restrict attention to the case when the susceptibility has only
a nonzero $zz$ component, because that greatly simplifies the form of the
Lorentz boosts.\footnote{This is not realistic, but the
presentation in this appendix is meant to be pedagogical.  The diligent reader
can make the generalization to the isotropic case.}
  We will work in a ($3+1$)-dimensional wave-vector
 representation.  Starting from Eq.~(\ref{Lorentzforce}), we have for the
force on the dielectric body in its rest frame
\be
F_z=\int(d\mathbf{r})\int\frac{d\omega\,d\nu}{(2\pi)^2}\int\frac{(d
\mathbf{k})}{(2\pi)^3}\frac{(d\mathbf{q})}{(2\pi)^3}e^{-i(\omega+\nu)t}
e^{i\mathbf{(k+q)\cdot r}}\,\mathbf{P}(\mathbf{k},\omega)iq_z\mathbf{E}(
\mathbf{q},\nu),
\ee
where all quantities are in the rest frame of the body.  Here,
we have omitted the first term in Eq.~(\ref{Lorentzforce}), because it will
always vanish {in view of the FDT constraint $\omega+\nu=0$.

\subsection{First-order friction}
To begin, we examine this in first order, where in place of Eqs.~(\ref{eppe}) 
we have in momentum space
(leaving off the component indices, which are always $z$ in our case)
\begin{subequations}
\bea
P(\mathbf{k},\omega)&=&\int\frac{(d\mathbf{k'})}{(2\pi)^3} \chi(\mathbf{k-k'})
E(\mathbf{k}',\omega),\\
\tilde{E}(\mathbf{k},\omega)&=&\Gamma({\mathbf{k},\omega)\tilde{P}
(\mathbf{k},\omega}).
\eea
\end{subequations}
(Quantities without tildes are in the rest frame of the body, while
those with tildes are in the rest frame of the radiation.)
Expanding $P$ to first order in $E$, we have
\be
F_z^{(1,0)}=\int\frac{d\omega}{2\pi}\frac{d\nu}{2\pi}
\frac{(d\mathbf{k})}{(2\pi)^3}
\frac{(d\mathbf{k'})}{(2\pi)^3}e^{-i(\omega+\nu)t}(-ik_z)\chi(\mathbf{k-k'},
\omega)E(\mathbf{k}',\omega)E(-\mathbf{k},\nu),\label{f10}
\ee
where the body rest frame value of the electric field, $E$ is
related to that in the radiation rest frame, $\tilde{E}$, by
\be
E(\mathbf{k},\omega)=\tilde{E}(\tilde{\mathbf{k}},\tilde\omega),
\ee
(this is where the simplification to have only $z$ polarization is crucial).
The boosted momentum and frequency variables are
\be
\tilde{k}_z=\gamma(k_z+\omega v), \quad \tilde{k}_{x,y}=k_{x,y},
\quad \tilde{\omega}=\gamma(\omega+k_z v),
\ee in terms of the relativistic dilation factor $\gamma=(1-v^2)^{-1/2}$.
Then, we use the FDT in the form
\be
\langle \mathcal{S} 
\tilde{E}(\tilde{\mathbf{k}},\tilde{\omega})\tilde{E}(\tilde{\mathbf{q}},
\tilde{\nu})\rangle=(2\pi)^4
\delta(\tilde\omega+\tilde\nu)\delta(\tilde{\mathbf{k}}+\tilde{\mathbf{q}})
\im \Gamma(\mathbf{\tilde{k}},\tilde{\omega})\coth\frac{\beta \tilde{\omega}}2.
\label{fdta}
\ee
The product of $\delta$ functions is invariant under a Lorentz boost,
\be
\delta(\tilde\omega+\tilde\nu)\delta(\tilde{\mathbf{k}}+\tilde{\mathbf{q}})=
\delta(\omega+\nu)\delta(\mathbf{k}+\mathbf{q}),
\ee
as is the $zz$ component of the vacuum Green's dyadic:
\be
\tilde{\Gamma}(\tilde{\mathbf{k}},\tilde{\omega})=
\Gamma(\mathbf{k},\omega)=\frac{k_\perp^2}{k^2-(\omega+i\epsilon)^2}.
\ee
The imaginary part of the latter is
\be
\im \Gamma(\mathbf{k},\omega)=k_\perp^2\pi\sgn(\omega)\delta(k^2-\omega^2).
\ee
Putting this into the formula for the force (\ref{f10}) we obtain
\bea
F^{(1,0)}_z&=&
\int\frac{d\omega(d\mathbf{k})}{(2\pi)^4}(-ik_z)\chi(\bm{0},\omega)
\im \Gamma(\mathbf{k},\omega)\coth\frac{\beta\tilde{\omega}}2\nn\\
&=&\frac1{(2\pi)^2}\int_0^\infty d\omega\int_{-\omega}^\omega d k_z
\im \chi(\bm{0},\omega)
k_x(\omega^2-k_z^2)\frac1{e^{\beta\gamma(\omega+k_z v)}-1}.\label{eff10}
\eea
Introducing the variable $y$ through $\gamma(\omega+k_z v)=\omega y$, we
obtain exactly the result (\ref{1stfric}) for the $ZZ$ polarization: there is
no contribution at $v=0$ and the PP or the (0,1) contribution is zero,
because the corresponding integral is odd in $k_z$.

\subsection{Second-order force and friction}
To obtain a propulsive force, we have to go to second order in the
susceptibility.  We begin by expanding the free energy so that
$P$ is expanded once, and $E$ is expanded twice, the term we refer to as
(1,2).  For simplicity of notation, we leave off integration elements:
\be
P(\mathbf{k},\omega) E(\mathbf{k}',\nu)=
\chi(\mathbf{k-k''},\omega)\tilde{E}(\tilde{\mathbf{k}}'',\tilde{\omega})
\Gamma(\mathbf{\tilde{k}'},\tilde{\nu})\chi(\mathbf{k'-k'''},\nu) 
\tilde{E}(\tilde{\mathbf{k}}''',\tilde{\nu}),
\ee
doing successive Lorentz boosts back and forth between the two natural
frames, and then use the FDT  (\ref{fdta}).
The result for this force component is
\be
F_z^{(1,2)}=\int\frac{d\omega\,(d\mathbf{k})\,(d\mathbf{k}')}{(2\pi)^7}
k'_z|\chi(\mathbf{k-k'},\omega)|^2\im \Gamma(\mathbf{k},\omega)\im\Gamma(
\mathbf{k'},\omega)\coth\frac{\beta\tilde\omega}2,\quad \tilde\omega=
\gamma(\omega+k_zv),\label{f12}
\ee
where the second imaginary part of a Green's dyadic comes from the
symmetry of the integrand under
 $(\omega,\mathbf{k,k'})\to -(\omega,\mathbf{k,k'})$,
and we have noted that
\be
\chi(\mathbf{k},\omega)^*=\chi(-\mathbf{k},-\omega),
\ee
because the underlying coordinate-space response function is real.

Following the same procedure, expanding $P$ out to third order, we
find
\be
F^{(3,0)}_z=\int\frac{d\omega\,(d\mathbf{k})\,(d\mathbf{k}')}{(2\pi)^7}
(-ik_z)\chi(\mathbf{k-k'},\omega)\chi(\mathbf{k'-k},\omega)
 \Gamma(\mathbf{k'},\omega)\im\Gamma(
\mathbf{k},\omega)\coth\frac{\beta\tilde\omega}2.\label{f30}
\ee
The two contributions to the EE force correspond to what
we called I [Eq.~(\ref{f30})] and II [Eq.~(\ref{f12})] in Ref.~\cite{guo}, 
We note that the (3,0) contribution vanishes at $v=0$ because it is then
antisymmetric under the reflection $(\mathbf{k,k'})\to-(\mathbf{k,k'})$.
It is possible then to show that the $v=0$ contribution of $F^{(1,2)}$
coincides with that found from Eq.~(\ref{FEEz}). namely,
\be
F^{\rm EE}_z\bigg|_{v=0}=
\int(d\mathbf{r})(d\mathbf{r'})\frac{d\omega}{2\pi}
\im[\chi(\mathbf{r};\omega)\chi(\mathbf{r}';\omega)^*]
\im\Gamma(\mathbf{r-r'};\omega)\nabla_z\im\Gamma(\mathbf{r-r'};\omega)
\coth\frac{\beta\omega}2,
\ee
having the same form as Eq.~(\ref{qvf}).

In general, however, because the product of susceptibilities in $F_z^{(3,0)}$
is not real, the real part of $\Gamma(\mathbf{k'},\omega)$ occurs, which 
leads to divergences when the integration over $\mathbf{k}'$ is performed.
The solution to this difficulty was provided in Ref.~\cite{guo}: when
$F_z^{(1,0)}$ [Eq.~(\ref{eff10})] and $F_z^{(3,0)}$ [Eq.~(\ref{f30})] are
combined, we have
\be
F_z^{(1,0)+(3,0)}=\int\frac{d\omega\,(d\mathbf{k})\,(d\mathbf{k}')}{(2\pi)^7}
(-ik_z)\hat{\chi}(\mathbf{k-k'},\omega)\im\Gamma(\mathbf{k},\omega)\coth
\frac{\beta\tilde{\omega}}2,
\ee
where
\be
\hat{\chi}(\mathbf{k-k'},\omega)=\chi(\mathbf{k-k'},\omega)\left[
(2\pi)^3\delta(\mathbf{k-k'})+\Gamma(\mathbf{k}',\omega)
\chi(\mathbf{k'-k},\omega)\right].
\ee
These are the first two terms of an infinite series, 
% [see Appendix \ref{gka}],
 written in symbolic notation:
\be
\hat\chi=\chi^{\vphantom{q}}_0\left(1-\Gamma\chi^{\vphantom{q}}_0\right)^{-1},
\ee
where we have denoted by $\chi^{\vphantom{q}}_0$, 
the bare intrinsic susceptibility, what we previously
called $\chi$.  Then, breaking up $\Gamma$ into its real and imaginary
parts, we have for the effective susceptibility, 
just as in Eq.~(4.31) of Ref.~\cite{guo},
\begin{subequations}
\be
\hat\chi=\chi(1-i\im\Gamma\chi)^{-1},
\ee
in terms of the renormalized intrinsic  susceptibility
\be
\chi=\chi^{\vphantom{q}}_0(1-\re \Gamma\chi^{\vphantom{q}}_0)^{-1}.
\ee
\end{subequations}
This is how the bare intrinsic susceptibilty is dressed by the radiation field.
(After all, the momentum-space Fourier transform of $\chi$ is the 
generalization of the polarizability discussed in Ref.~\cite{guo}.)
Therefore, we conclude that the renormalized version of $F^{(3,0)}$ should
read
\be
F^{(3,0)}_{z,R}=\int\frac{d\omega\,(d\mathbf{k})\,(d\mathbf{k}')}{(2\pi)^7}
k_z\re[\chi(\mathbf{k-k'},\omega)\chi(\mathbf{k'-k},\omega)]
\im \Gamma(\mathbf{k'},\omega)\im\Gamma(
\mathbf{k},\omega)\coth\frac{\beta\tilde{\omega}}2,\label{f30r}
\ee
which is free of all divergences 
because the real part
of $\bm{\Gamma}$ has been absorbed into the redefinition of $\chi$.

For the PP fluctuation contributions, the calculation proceeds similarly.
Now we either expand the free energy out to third order in $E$, giving
a (0,3) contribution, or expand  $P$ out to second order and
$E$ out to first, the (2,1) term.  The FDT
(\ref{ppfdt}),  in the Fourier domain, reads
\begin{subequations}
\be
\langle \mathcal{S} P(\mathbf{k},\omega)P(\mathbf{k'},\nu)\rangle
=2\pi\delta(\omega+\nu)
\Im\chi(\mathbf{k+k'},\omega)\coth\frac{\beta'\omega}2,
\ee
with
\be
\Im\chi(\mathbf{k+k'},\omega)=\frac1{2i}\left[\chi(\mathbf{k+k'},\omega)
-\chi(\mathbf{k+k'},-\omega)\right],
\ee
\end{subequations}
neither the imaginary nor the anti-Hermitian part.  The two contributions
to the force  are, respectively,\footnote{Note that no renormalization of
$F_z^{(0,3)}$ is necessary, because $F_z^{(0,1)}$ vanishes.}
\begin{subequations}
\bea
F_z^{(0,3)}&=&i\int\frac{d\omega(d\mathbf{k})(d\mathbf{k'})}{(2\pi)^7} k_z
\Gamma(\mathbf{k},-\omega)\Gamma(\mathbf{k}',-\omega)
\chi(\mathbf{k-k'},-\omega)
\Im\chi(\mathbf{k'-k},-\omega)\coth\frac{\beta'\omega}2, \label{f03}
\\
F_z^{(2,1)}&=&-i\int\frac{d\omega(d\mathbf{k})(d\mathbf{k'})}{(2\pi)^7} k_z
\Gamma(\mathbf{k},-\omega)\Gamma(\mathbf{k'},\omega)\chi(\mathbf{k-k'},\omega)
\Im\chi(\mathbf{k'-k},\omega)\coth\frac{\beta'\omega}2.\label{f21}
\eea
\end{subequations}
Using the symmetries $(\mathbf{k,k'})\to-(\mathbf{k,k'})$ and 
$\mathbf{k\leftrightarrow k'}$, 
we obtain the total PP contribution to the force
\be
F^{(0,3)+(2,1)}=-\frac14\int \frac{d\omega(d\mathbf{k})(d\mathbf{k'})}
{(2\pi)^7}(k_z-k_z')\im\Gamma(\mathbf{k},\omega)\im\Gamma(\mathbf{k'},\omega)
\left[|\chi(\mathbf{k'-k},\omega)|^2-|\chi(\mathbf{k-k'},\omega)|^2
\right]\coth\frac{\beta'\omega}2.
\ee
and the total $v=0$ propulsive force is
\bea
F^{EE+PP}&=&-\int_0^\infty\frac{d\omega}{2\pi}\int 
\frac{(d\mathbf{k})(d\mathbf{k'})}{(2\pi)^6}(k_z-k_z')\im\Gamma(\mathbf{k},
\omega)\im\Gamma(\mathbf{k'},\omega)
\left[|\chi(\mathbf{k-k'},\omega)|^2-|\chi(\mathbf{k'-k},\omega)|^2
\right]\nn\\
&&\qquad\times
\left[\frac1{e^{\beta\omega}-1}-\frac1{e^{\beta'\omega}-1}\right].
\eea
This is simply the Fourier representation of Eq.~(\ref{qvf})
for the propulsive force.
%This agrees with the propulsive force given in Sec.~\ref{sec1}.

As for the $v\ne0$ frictional force, 
which arises only from the EE fluctuations, 
we can compare with the 
second-order results of Ref.~\cite{guo}, where we considered a point particle
with a real polarizability.  Then there are only EE fluctuations, and
the spatial Fourier transform of the susceptibility is the polarizability
of the particle, $\chi(\mathbf{k-k'},\omega)=\alpha(\omega)$.  
It is straightforward to show that the EE force is entirely
frictional and coincides with Eq.~(3.16) of Ref.~\cite{guo}.  This is
because
\be
\int dk_x \im \Gamma(\mathbf{k},\omega)=\int dk_x (k_x^2+k_y^2)\im
\frac1{k_x^2+k_y^2+k_z^2-(\omega-i\epsilon)^2}=\pi \sgn(\omega)
\frac{\omega^2-k_z^2}{\sqrt{\omega^2-k_y^2-k_z^2}}\theta(\omega^2-k_y^2-k_z^2).
\ee
That is, the frictional force in this case is
\be
F^{(3,0)+(1,2)}=\frac1{32\pi^3}\int_0^\infty d\omega\int_{-\omega}^\omega dk_z
\int_{-\omega}^\omega dk_{z}'\, \alpha^2(\omega)(k_z+k_z')
(\omega^2-k_z^2)
(\omega^2-k_z^{\prime 2})\frac1{e^{\beta\gamma(\omega+k_zv)}-1}.
\label{frictions}
\ee
The $F^{(1.2)}$ contribution here, proportional to $k'_z,$ vanishes because
it is odd in $k'_z$. 
%If $F^{(1,2)}$ is expanded to first order in $v$, the
%frictional force so arising, proportional to $k_zk_z'$, vanishes by
%$k_z\leftrightarrow k'_z$ symmetry.
%Evidently, the $F^{(1,2)}$ term, proportional to $k_z'$, vanishes by symmetry.
%This remains true for any object with uniform susceptibility. 
This statement
is true for any object with uniform susceptibility; that is, the frictional
force only arises from $F^{(3,0)}$, while the propulsive force on a 
nonuniform object has an EE contribution coming only from $F^{(1,2)}$.
(We already recognized the latter in Sec.~\ref{sec1} (see below 
Eq.~(\ref{secordsus})) 
while the former was observed in Ref.~\cite{guo}.)

For a small object, with polarizability proportional to its volume, this 
second-order friction would be expected to be much smaller than the 
first-order friction given in Eq.~(\ref{1stfric}).
Consider the thin needle (half length $b$)
as an example.  The friction comes only from 
$F^{(3,0)}$, and, in the approximation 
that the friction arises from a region where
the susceptibility is real and uniform, is 
the Einstein-Hopf (low velocity) limit,
\be
F^{(2)}_f=
 \frac4{\pi^3} C^2 v T^6\int_0^\infty dz \,z^6\chi(2T z)^2\frac{r(z b T)}{
\sinh^2 z},\label{forcefn}
\ee
where
\be r(t)=\int_{-1}^1 dx\int_{-1}^1 dy \frac{x^2(1-x^2)(1-y^2)}{(x-y)^2}\sin^2[(x-y)
t].
\ee
In Eq.~(\ref{forcefn}) we use
Eq.~(\ref{f30r}) rather than Eq.~(\ref{frictions}), because 
we consider a long needle. 
The integral $r(t)$ may be given in closed form in terms of the sine-integral
function. but only the large-$t$ behavior is relevant for
a 1 cm object at room temperature because typical values of $t$ are of order
$bT=1250$ at room temperature:
\be
r(t)\sim \frac{16\pi t}{105},\quad t\gg1.
\ee
For a representative value of susceptibility $\chi_0$, 
and a gold needle of radius 50 nm and length $b=1$ cm,
the force (\ref{forcefn}) is smaller than
the frictional force found in Eq.~(\ref{fpsf}) by a factor of
$5\times 10^{-8} \chi_0^2$.

\section{Radiated energy}
\label{appb}
In Ref.~\cite{guo2} we derived the energy lost by a polarizable  particle out
of thermal equilibrium by looking at the rate with which the particle does
work on the field.  Alternatively, we can examine the rate at which the 
energy flows across a surface surrounding a body, which follows from 
energy conservation,
\be
\frac\partial{\partial t}u+\bm{\nabla}\cdot \mathbf{S}+\mathbf{j\cdot E}=0,
\ee
provided, as is the case here, that the situation is stationary.  Here
the energy flux vector or Poynting vecctor is
\be
\mathbf{S(r,t)}=\int\frac{d\omega}{2\pi}\frac{d\nu}{2\pi}e^{-i(\omega+\nu)
t}\mathbf{E(r;\omega)}\times\left[ \frac1{i\nu}\bm{\nabla}
\times\mathbf{E(r;\nu)}\right].
\ee
If the electric field is expanded out to first order with the polarization
as given in Eq.~(\ref{EofP}), and we use the FDT (\ref{ppfdt}) for the 
polarization
field product, we find for the PP fluctuation part of the Poynting vector
\be
S^{\rm PP}_k(\mathbf{r})=-\int(d\mathbf{r'})\frac{d\omega}{2\pi}\frac1{i\omega}
\Gamma_{ij}(\mathbf{r-r'};\omega)\im \chi_{jl}^{\vphantom{q}}
(\mathbf{r'};\omega)\coth
\frac{\beta'\omega}2\left[\nabla_k \Gamma_{il}(\mathbf{r-r'};-\omega)
-\nabla_i\Gamma_{kl}(\mathbf{r-r'};-\omega)\right].\label{SPP}
\ee
Now we imagine surrounding the body with a  sphere of radius $R$, large
compared to the size of the body and centered on it. 
 Then the Green's dyadic appearing here is
asymptotically [see Eq.~(\ref{gammaprime})]
\be
\bm{\Gamma}(\mathbf{R};\omega)\sim \omega^2(\bm{1}-\hat{\mathbf{R}}\hat{
\mathbf{R}}) \frac{e^{i\omega R}}{4\pi R},\quad R\gg a,
\ee
where $a$ is a typical size of the object.  The corresponding flux vector is
\begin{subequations}
\bea
S^{\rm PP}_k(\mathbf{r})
&=&\int\frac{d\omega}{2\pi}(d\mathbf{r'})\frac{\omega^4}
{(4\pi R)^2}\hat{R}_k(\delta_{ij}-\hat{R}_i\hat{R}_j)
\im\chi_{ij}^{\vphantom{q}}(\mathbf{r'};
\omega)\coth\frac{\beta'\omega}2\\ 
&\to& 2\int\frac{d\omega}{2\pi}(d\mathbf{r'}) \frac{\omega^4}{(4\pi R)^2}
\hat{R}_k\im\chi(\mathbf{r'};\omega)\coth\frac{\beta'\omega}2,
\eea
\end{subequations}
where the second form applies for isotropic susceptibility.
Integrating this over a large sphere surrounding the body gives the total
energy  emitted per unit time:
\be
\int d\sigma \hat{\mathbf{R}}\cdot \mathbf{S}^{\rm PP}(\mathbf{r})
=\frac1{4\pi^2}\int d\omega\,\omega^4\im\alpha(\omega)\coth
\frac{\beta'\omega}2.
\ee
Here, we use the lowest-order susceptibility, $\alpha(\omega)
=\int(d\mathbf{r'})\chi(\mathbf{r'};\omega)$.  
This total power agrees
with Eq.~(2.21) of Ref.~\cite{guo2}.

What we did in the preceding paragraph is what we denote as the (1,1) term in
the expansion.  To obtain the EE fluctuation contribution, we compute, in the
same way, the (2,0) and (0,2) expansions.  The result is as expected; the total
flux vector for large $R$, $R\gg r'$  being for arbitrary polarizability
\be
S_k=S_k^{\rm PP}+S_k^{\rm EE}
 =\frac{\hat{R}_k}{(4\pi R)^2}\int\frac{d\omega}{2\pi}\omega^4(\delta_{ij}-
\hat{R}_i\hat{R}_j)\im\alpha_{ij}(\omega)\left(\coth\frac{\beta'\omega}2-
\coth\frac{\beta\omega}2\right).
\ee

\section{Dielectric in front of a mirror}
\label{appc}
To understand the sense of the force, which in all the examples we have 
considered is in the direction of the metal side of the composite object,
we consider a simple example of a perfectly conducting plane, a perfect mirror,
located at $z=0$, attached to which is a dilute dielctric body, as shown in
Fig.~\ref{fig:mirdiel}.
\begin{figure}
\includegraphics{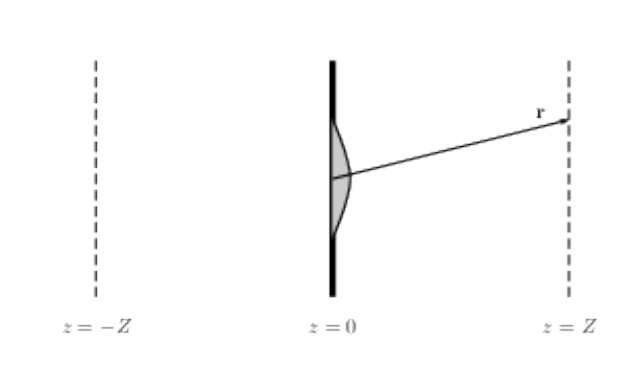}
\caption{Perfectly reflecting mirror located in the plane $z=0$
to which a dilute dielectric is attached.
The radiation fields are observed on the distant screen at $z=Z$.  The
screen to the left of the mirror sees nothing but the background field as
modified by the mirror.  The dielectric is at temperature $T'$ while the
background has temperature $T$.}
\label{fig:mirdiel}
\end{figure}
It is evident from the diagram that, because energy is radiated to the right
by the composite assembly, there will be a force on that assembly to the left,
in the $-z$ direction.  The purpose of this appendix is to make that picture
quantitative.

We treat the effect of the dielectric perturbatively, to first order in
its susceptibility, $\chi(\mathbf{r'};\omega)$.  The effect of the mirror
is to replace the free electromagnetic Green's dyadic $\bm{\Gamma}^{(0)}
(\mathbf{r-r'};\omega)$ by
the Green's dyadic \cite{LS}
\be
\bm{\Gamma}(\mathbf{r,r'};\omega)=\bm{\Gamma}^{(0)}(\mathbf{r-r'};\omega)-
\bm{\Gamma}^{(0)}(\mathbf{r-r'}+2z'\mathbf{\hat{z};\omega})
\cdot(\bm{1}-2\mathbf{
\hat z\hat z}),
\ee
which has an obvious interpretation in terms of an image charge.  We are 
interested in fields far from the mirror/dielectric assembly, $r\gg r'$,
where $\mathbf{r'}$ locates a point within the dielectric.
In the radiation zone, then,
\be
\bm{\Gamma}(\mathbf{r,r'};\omega)
\sim 2 \bm{\Gamma}^{(0)}(\mathbf{r};\omega)\cdot \mathbf{
\hat z\hat z}.
\ee
For large $r$, the only surviving components of the Green's dyadic are
from Eq.~(\ref{gammaprime}),
\be
\Gamma_{xz}\sim -\frac{2\omega^2}{4\pi r}\frac{xz}{r^2}e^{i\omega r},
\quad
\Gamma_{yz}\sim -\frac{2\omega^2}{4\pi r}\frac{yz}{r^2}e^{i\omega r},
\quad
\Gamma_{zz}\sim \frac{2\omega^2}{4\pi r}\left(1-\frac{z^2}{r^2}
\right)e^{i\omega r}.
\ee

Now we calculate the first-order energy flux as in Appendix \ref{appb}, see
Eq.~(\ref{SPP}):
\be
S_z=\int\frac{d\omega}{2\pi}(d\mathbf{r'})\frac1\omega\im\chi(\mathbf{r'};
\omega)\left(\coth\frac{\beta\omega}2-\coth\frac{\beta'\omega}2\right)
\im\left\{\Gamma_{ij}(\mathbf{r,r'};\omega)\left[\partial_z\Gamma_{ij}(\mathbf{
r,r'};-\omega)-\partial_i\Gamma_{zj}(\mathbf{r,r'};-\omega)\right]\right\},
\ee
for isotropic polarizability.
When the asymptotic forms of the Green's dyadic are inserted into this, we find
\be
S_z\sim-\frac1{4\pi^2}\frac{z\rho^2}{(\rho^2+z^2)^{5/2}}\int \frac{d\omega}{2\pi} \omega^4
\im\alpha(\omega)\left(\coth\frac{\beta\omega}2-\coth
\frac{\beta'\omega}2\right),
\ee
where $\rho=\sqrt{x^2+y^2}$ is the radial distance on the screen.
Here we have introduced again $\alpha(\omega)=\int (d\mathbf{r'}) \chi(\mathbf{
r'};\omega)$. 
Now we integrate $S_z$ over the infinite screen
%at $z=Z$ by introducing polar coordinates on that plane:
%\be
%r=Z\sec\theta,\quad \rho=\sqrt{x^2+y^2}=Z \tan\theta.
%\ee
%Therefore, 
and obtain the total power radiated:
\be
P=2\pi\int_0^\infty \rho\,d\rho\, S_z=
-\frac2{3\pi^2}\int_0^\infty d\omega\,\omega^4 \im\alpha(\omega)\left(
\frac1{e^{\beta\omega}-1}-\frac1{e^{\beta'\omega}-1}\right).\label{pmd}
\ee
The sign means that, if $T'>T$, the energy flux is, as is obvious, in the $+z$
direction.

For the force of the composite mirror/dielectric object, we compute the
surface integral of the $zz$ component of the electromagnetic stress tensor,
$T_{zz}$, over the same surface, as follows from the local
statement of momentum conservation,
\be
\frac{\partial}{\partial t}\mathbf{G}+\bm{\nabla}\cdot\mathbf{T}+\mathbf{f}=
\bm{0},
\ee
where $\mathbf{f}$ is the force density on all the free and bound charges, 
$\mathbf{T}$
is the stress tensor, and $\mathbf{G}$ is the momentum density of the field.
[The contributions to $S_z$ and $T_{zz}$ from the mirror alone in vacuum is
 zero.]  The force is, in our static situation,
\be
F_z=-\int_{z=Z}  d\sigma\,T_{zz}
=-\frac12\int_{z=Z} d\sigma\left(E_x^2+E_y^2-E_z^2+H_x^2+H_y^2-H_z^2\right).
\ee
The FDT gives PP contributions [that we referred to as (1,1) expansion terms]
and EE contributions [(2,0) and (0,2) contributions].  A straightforward
calculation for the former leads to the following radiation-zone result:

\be
F^{\rm PP}_z=-\int_{z=Z} \rho\,d\rho \,2\pi \frac{z^2\rho^2}{(\rho^2+z^2)^3}
\int\frac{d\omega}{2\pi}\im\alpha(\omega)\frac{\omega^4}{4\pi^2}
\coth\frac{\beta'\omega}2,
\ee
which exhibits the same positional dependence on the screen as the energy flux.
The corresponding calculation for the EE contributions involves recognizing
the  average
\be
\im \left(e^{i\omega R}\right) e^{-i \omega R}\to -\frac{i}2,\quad
\im\left( i\, e^{i\omega R}\right)i\,e^{i \omega R}\to +\frac{i}2;
\ee
the result of including the EE and PP contributions is 
\be
F_z=\frac1{8\pi^2}\int_0^\infty d\omega\,\omega^4\im\alpha(\omega)\left(\coth
\frac{\beta\omega}2-\coth\frac{\beta'\omega}2\right).\label{fzmd}
\ee
Indeed, the force is directed in the $-z$ direction, but the precise
connection between the energy radiated (\ref{pmd})
 and the force on the composite object (\ref{fzmd}) is obscure.
%\textcolor{red}{Evidently, the principle of virtual work does not apply
%in this situation.}


\begin{thebibliography}{99}

\bibitem{EinsteinHopf}
A. Einstein and L. Hopf, Statistische Untersuchung der Bewegung eines
Resonators in einem Strahlungsfeld, Ann. Phys.
(Leipzig) {\bf 338}, 1105 (1910).

\bibitem{Mrkt}
V. Mkrtchian, V. A. Parsegian, R. Podgornik, and W. M. Saslow,
Universal thermal radiation drag on neutral objects, Phys.\ Rev.\ Lett.\
{\bf91}, 220801 (2003). 

\bibitem{dedkov2}
G. V. Dedkov and A. A. Kyasov, Tangential force and
heating rate of a neutral relativistic particle mediated by
equilibrium background radiation, Nucl.\ Instrum.\ Methods
Phys.\ Res., Sect. B {\bf268}, 599 (2010).

\bibitem{Lach}
G. {\L}ach, M. DeKieviet, and U. D.Jentschura, Einstein-Hopf drag,
Doppler shift of thermal radiation and blackbody drag: 
Three perspectives on quantum friction, Cent.\ Eur.\ J.\ Phys.\
{\bf10}, 763(2012).



\bibitem{pieplow}
 G. Pieplow and C. Henkel, Fully covariant radiation force
on a polarizable particle, New J. Phys.\ {\bf15}, 023027 (2013).

\bibitem{Volokitin0} A. I. Volokitin, Friction force at the motion
of a small relativistic neutral particle with respect to blackbody radiation,
JETP Letters, {\bf 101}, 427--433 (2015).

\bibitem{vp}
A. I. Volokitin and B. N. J. Persson, {\it Electromagnetic Fluctuations at the 
Nanoscale} (Springer, Berlin, 2017).



\bibitem{teod}
E. V. Teodorovich,  Contribution of macroscopic van der Waals interactions to
frictional force. Proc. R. Soc.\ Lond.\ A  {\bf 362}, 71--77 (1978).

\bibitem{lev}
L. S. Levitov, Van der Waals friction. Europhys.\ Lett. {\bf8}, 499--504
(1989).

\bibitem{pendry} J. B. Pendry, Shearing the vacuum---quantum friction,
J. Phys.\ Condens.\ Matter {\bf9}, 10301 (1997).

\bibitem{volokitin} A. I. Volokitin and B. N. J. Persson,
Theory of friction: The
contribution from a fluctuating electromagnetic field,
J. Phys.\ Condens.\ Matter {\bf11}, 345 (1999).

\bibitem{dedkov} G. V. Dedkov and A. A. Kyasov, The relativistic theory of
fluctuation electromagnetic interactions of moving neutral
particles with a flat surface,
Phys.\ Solid State {\bf45}, 1815 (2003).

\bibitem{davies} P. C. W. Davies, Quantum vacuum friction, J. Opt.\ B {\bf7},
S40 (2005).

\bibitem{barton} G. Barton, On van der Waals friction. II: Between atom and
half-space, New J. Phys.\ {\bf12}, 113045 (2010).

\bibitem{Zhao} R. Zhao, A. Manjavacas, F. J. G. de Abajo, and J. B. Pendry,
  Rotational quantum friction, Phys.\ Rev.\ Lett.\ {\bf109}, 123604 (2012).

\bibitem{Silveirinha} M. G. Silveirinha, Theory of quantum friction,
New J. Phys.\ {\bf16}, 063011 (2014).

\bibitem{Pieplow15} G. Pieplow and C. Henkel, Cherenkov friction on a
neutral particle moving parallel to a dielectric, J. Phys.\ Condens. Matter
{\bf 27}, 214001 (2015).

\bibitem{intravaia} F. Intravaia, V. E. Mkrtchian, S. Y. Buhmann, S. Scheel,
D. A. R. Dalvit, and C. Henkel, Friction forces on atoms
after acceleration, J. Phys.\ Condens.\ Matter {\bf27}, 214020 (2015).

\bibitem{rev} J. S. H{\o}ye, I. Brevik and K. A. Milton,
The reality of Casimir friction  [arXiv:1508.00626], Symmetry {\bf 8} (5), 
29 (2016), doi:10.3390/sym8050029.

\bibitem{pan}
D. Pan, H. Xu, and F. J. Garc\'{i}a de Abajo, Magnetically
activated rotational vacuum friction, Phys.\ Rev.\ A {\bf 99}, 062509
(2019).

\bibitem{Dedkov20} G. V. Dedkov and A. A. Kyasov, Nonlocal friction forces
in the particle-plate and plate-plate configurations: Nonretarded 
approximation, Surface Sci. {\bf 700}, 121681 (2020).


\bibitem{Farias} 
M. B. Far\'{i}as, F. C. Lombardo, A. Soba, P. L. Villar,
and R. Decca, 
Towards detecting traces of non-contact quantum friction in the corrections 
of the accumulated geometric phase,
npj Quantun Information  {\bf 6}, 25 (2020).
 https://doi.org/10.1038/s41534-020-0252-x


\bibitem{Dedkov22} G. V. Dedkov,
Nonequilibrium Casimir-Lifshitz force and anomalous
radiation heating of a small particle, arXiv:2207.13769.


\bibitem{marty}
M. Oelschl\"{a}ger, Fluctuation-induced phenomena in nanophotonic systems,
Ph.D. thesis, Institut f\"ur Physik, Humboldt  Universit\"at zu Berlin, 2020.

\bibitem{Wang} T.-B. Wang, Y. Zhou, H.-Q. Mu, K. Shehzad, D.-J. Zhang,
W.-X. Liu, T.-B. Yu, and Q.-H. Liao, Enhancement of lateral Casimir
force on a rotating body near a hyperbolic material, Nanotechnology
{\bf 33}, 245001 (2022).

\bibitem{Brevik22} I. Brevik, B. Shapiro, and M. G. Silveirinha,
Fluctuational electrodynamics in and out of equilibrium,
Int.\ J. Mod.\ Phys.\ A {\bf 37}, 2241012 (2022).




\bibitem{guo}X. Guo, K. A. Milton, G. Kennedy, W. P. McNulty,
N. Pourtolami, and Y. Li,
Energetics of quantum vacuum friction: Field fluctuations,
[arXiv:2108.01539] Phys.\ Rev.\ D {\bf104}, 116006 (2021).

\bibitem{guo2}
X. Guo, K. A. Milton, G. Kennedy, W. P. McNulty,
N. Pourtolami, and Y. Li,
Energetics of quantum vacuum friction. II: Dipole fluctuations and field
fluctuations, [arXiv:2204.10886], Phys.\ Rev.\ D {\bf106}, 016008 (2022).


%\bibitem{EQVFII} X. Guo, K. A. Milton, G. Kennedy, W. P. McNulty, N. 
%Pourtolami, and Y. Li, ``Energetics of quantum vacuum friction. II. Dipole
%fluctuations and field fluctuations,'' Phys.\ Rev.\ D {\bf 106}, 016008 (2022).








\bibitem{reiche}
D. Reiche, F. Intravaia, J.-T. Hsiang, K. Busch, and B.-L.
Hu, Nonequilibrium thermodynamics of quantum friction,
Phys.\ Rev.\ A {\bf102}, 050203(R) (2020).






\bibitem{krugeremig} M. Kr\"{u}ger, T. Emig, and M. Kardar,
Nonequilibrium Electromagnetic Fluctuations: Heat transfer and
interactions, Phys.\ Rev.\ Lett.\ {\bf 106}, 210404 (2011).

\bibitem{Ott}
A. Ott, P. Ben-Abdallah, and S.-A. Biehs, Circular heat and momentum
flux radiated by magneto-optical nanoparticles, Phys.\ Rev.\ B {\bf 97},
205414 (2018).

\bibitem{maghrebi} M. F. Maghrebi, A. V. Gorshkov, and J. D. Sou,
Fluctuation-induced torque on a topological insulator out of thermal
equilibrium, Phys.\ Rev.\ Lett.\ {\bf 123}, 055901 (2019).

\bibitem{Khandekar}
C. Khandekar and Z. Jacob, Thermal spin photonics in the near-field
of nonreciprocal media,
New J. Phys.\ {\bf 21}, 103030 (2019)


\bibitem{khandekar0}
C. Khandekar, S. Buddhiraju, P. R.
Wilkinson, J. K. Gimzewski, A. W. Rodriguez, C. Chase,
S. Fan, %Nonequilibrium Casimir effects of nonreciprocal surface waves,'' 
Nonequilibrium lateral force and torque by thermally excited nonreciprocal
surface electromagnetic waves,
Phys.\ Rev.\ B {\bf 104}, 245433 (2021).

\bibitem{fogedby}
H. C. Fogedby and A. Imparato, Autonomous quantum rotator, EPL {\bf122},
10006 (2018).



\bibitem{guofan} C. Guo and S. Fan, Theoretical constraints
on reciprocal and non-reciprocal many-body radiative heat transfer,
[arXiv:2007.13274] Phys.\ Rev.\ B {\bf 102}, 085401 (2020).

\bibitem{fan} Y. Guo and S. Fan, A single gyrotropic particle
as a heat engine, [arXiv:2007.11234] ACS Photonics {\bf 8}, 1623--1629
(2021).

\bibitem{gelb} D. Gelbwaser-Klimovsky, N. Graham, M. Kardar,
and M. Kr\"{u}ger, Near field propulsion forces from nonreciprocal
media, Phys.\ Rev.\ Lett.\ {\bf 126}, 170401 (2021).


\bibitem{strekha}  B. Strekha, S. Molesky, P. Chao,
M. Kr\"uger, and A. Rodriguez,
Trace expressions and associated limits for non-equilibrium Casimir
torque,
[arXiv:2207.13646] Phys.\ Rev.\ A {\bf106}, 042222 (2022).

\bibitem{Ge} L. Ge, Negative vacuum friction in terahertz gain systems,
Phys.\ Rev.\ B {\bf 108}, 045406 (2023).

\bibitem{tutorial}
V. S. Asadchy, M. S. Mirmoosa, A. Diaz-Rubio, S. Fan, and S. A. Tretyakov,
Tutorial on electromagnetic nonreciprocity and its origins, Proc. IEEE,
{\bf 108}, (10) 1684--1727 (2020) [arXiv:2001.04848].

\bibitem{QVT} K. A. Milton. X. Guo, G. Kennedy, N. Pourtolami, and D. M DelCol,
Vacuum torque, propulsive forces, and anomalous tangential forces:
Effects of nonreciprocal media out of thermal equilibrium,
[arXiv:2306.02197]
Phys.\ Rev.\ A {\bf 108}, 022809 (2023). 

\bibitem{kennedy} G. Kennedy, Quantum torque on a non-reciprocal body out
of thermal equilibrium and induced by a magnetic field of arbitrary
strength, Eur.\ Phys.\ J. Spec.\ Top.\ {\bf 232}, 3197--3208 (2023). 
https://doi.org/10.1140/epjs/s11734-023-01068-0

\bibitem{Reid}
M. T. H. Reid, O. D. Miller, A. G. Polimeridis, A. W. Rodriguez,
E. M. Tomlinson, and S. G. Johnson,  Photon torpedoes and Rytov
pinwheels: Integral-equation modeling of non-equilibrium fluctuation-induced
forces and torques on nanoparticles, arXiv:1708.01985 (2017).

\bibitem{muller}
B. M\"{u}ller and M. Kr\"{u}ger, Anisotropic
particles near surfaces: Propulsion force and friction, Phys.\ Rev.\
A {\bf 93}, 032511 (2016).

\bibitem{Lambrecht}
A. Lambrecht and S. Reynaud,
Casimir force between metallic mirrors,
Eur. Phys. J. D {\bf 8}, 309--318 (2000)
https://doi.org/10.1007/s100530050041

\bibitem{ce} K. A. Milton and J. Schwinger,  
{\it Classical Electrodynamics}, 2nd ed.,  (CRC Press/Taylor and
Francis, Boca Raton, 2024).


\bibitem{synthesis}
M. Damani, N. Desai, B. P. Singh, R. S. Ningthoujam,
M Momin, and T. Khan, Synthesis of hollow gold nanoparticles: Impact
of variables on process optimization, J. Pharma. Sciences {\bf 111}, 2907-2916
(2022), https://doi.org/10.1016/j.xphs.2022.08.003

\bibitem{manjavacas}
J. R. Deop-Ruano, F. J. Garc\'{i}a de Abajo, and A. Manjavacas,
Thermal radiation forces on planar structures
with asymmetric optical response, Nanophotonics, 2024,
https://doi.org/10.1515/nanoph-2024-0121.

\bibitem{thruster}
%https://handwiki.org/wiki/Physics:Quantum${}_{_}$vacuum${}_{_}$thruster
Quantum Vacuum Thruster, Scholarly Community Encyclopedia,
https://encyclopedia.pub/entry/28846

\bibitem{Parsegian}V. A. Parsegian, {\it Van der Waals Forces: A Handbook
for Biologists, Chemists, Engineers, and Physicists\/} (Cambridge University
Press, 2006), p.~268.





%\bibitem{entropy}
%Y. Li, K. A. Milton, P. Parashar, G. Kennedy, N. Pourtolami, and X. Guo,
%Casimir self-entropy of nanoparticles with classical polarizabilities:        
%Electromagnetic field fluctuations, Phys.\ Rev.\ D {\bf 106}, 036002 (2022).










%\bibitem{Maghrebi}                                                             
%M. F. Maghrebi, A. V. Gorshkov, and J. D. Sau,                                 
%``Fluctuation-induced torque on a topological insulator out of thermal         
%equilibrium,'' Phys.\ Rev.\ Lett.\ {\bf 123}, 055901 (2019)                    






%\bibitem{LL} L. D. Landau and E. M. Lifshitz, {\it Statistical Physics, Part 1%}
% (Pergamon, Oxford, 1980), 3rd edition, Sec.~66.

%\bibitem{sse}
%K. A. Milton, P. Kalauni, P. Parashar, and Y. Li,
%Casimir self-entropy of a spherical electromagnetic $\delta$-function
%shell, [arXiv: 1707.09840], Phys.\ Rev.\ D {\bf96}, 085007 (2017).

%\bibitem{Pal}P. Palffy-Muhoray and D. A. Balzarini,
%The Clausius-Mossotti relation for anisotropic molecular fluids,
%Can.\ J. Phys.\ {\bf 59}, 375 (1981).

%\bibitem{Puska} M. J. Puska, R. M. Nieminen, and M. Manninen,
%Electronic polarizability of small metal spheres, Phys.\ Rev.\ B
%{\bf 31}, 3486 (1985).

%\bibitem{Kheirandish} A. Kheirandish, N. S. Javan, and H. Mohammadzedah,
%Modified Drude model for small gold nanoparticles surface plasmon
%resonance based on the role of classical confinement, Sci.\ Rep.\
%(2000) 10:6517 [https://doi.org/10.1038/s41598-020-63066-9].

\bibitem{LS} H. Levine and J. Schwinger, On the theory of electromagnetic
wave diffraction by an aperture in an infinite plane conducting screen,
Comm.\ Pure Appl.\ Math.\ III, {\bf4}, 355 (1950), reprinted in K. Milton
and J. Schwinger, {\it Electromagnetic Radiation: Variational Methods, 
Waveguides, and Accelerators} (Springer, Berlin, 2006).

\end{thebibliography}
\end{document}